\documentclass[conference]{IEEEtran}
\IEEEoverridecommandlockouts
\usepackage{amsmath,amssymb,amsfonts}
\usepackage{multirow}%
\usepackage{algorithm}%
\usepackage{algorithmicx}%
\usepackage{algpseudocode}%
\usepackage{array}
\usepackage[caption=false,font=footnotesize,labelfont=rm,textfont=rm]{subfig}
\usepackage{textcomp}
\usepackage{stfloats}
\usepackage{url}
\usepackage{verbatim}
\usepackage{graphicx}
\usepackage{listings}%
\usepackage{cite}
\usepackage{booktabs}
\usepackage{subfig}
\usepackage{threeparttable}
\usepackage{makecell}
\usepackage{doi}
\usepackage{xcolor}
\usepackage{tikz}

\def\BibTeX{{\rm B\kern-.05em{\sc i\kern-.025em b}\kern-.08em
    T\kern-.1667em\lower.7ex\hbox{E}\kern-.125emX}}

\begin{document}

\title{Specific Emitter Identification Handling Modulation Variation with Margin Disparity Discrepancy\\
\thanks{Xuanpeng Li is the corresponding author.\\
\\
This work has been submitted to the IEEE for possible publication. Copyright may be transferred without notice, after which this version may no longer be accessible.}
}

\author{
	\IEEEauthorblockN{
        Yezhuo Zhang, 
		Zinan Zhou, 
		Xuanpeng Li} 
	\IEEEauthorblockA{School of Instrument Science and Engineering, Southeast University, Nanjing, China\\
    Email: \{zhang\_yezhuo, zhouzinan919, li\_xuanpeng\}@seu.edu.cn}
} 
\maketitle

\begin{abstract}
In the domain of Specific Emitter Identification (SEI), it is recognized that transmitters can be distinguished through the impairments of their radio frequency front-end, commonly referred to as Radio Frequency Fingerprint (RFF) features. However, modulation schemes can be deliberately coupled into signal-level data to confound RFF information, often resulting in high susceptibility to failure in SEI. In this paper, we propose a domain-invariant feature oriented Margin Disparity Discrepancy (MDD) approach to enhance SEI's robustness in rapidly modulation-varying environments. First, we establish an upper bound for the difference between modulation domains and define the loss function accordingly. Then, we design an adversarial network framework incorporating MDD to align variable modulation features. Finally, We conducted experiments utilizing 7 HackRF-One transmitters, emitting 11 types of signals with analog and digital modulations. Numerical results indicate that our approach achieves an average improvement of over 20\% in accuracy compared to classical SEI methods and outperforms other UDA techniques. Codes are available at https://github.com/ZhangYezhuo/MDD-SEI.
\end{abstract}


\section{Introduction}\label{sec1}

Specific Emitter Identification (SEI) entails distinguishing individual radiation sources by comparing the Radio Frequency Fingerprint (RFF) features, which arise from hardware differences unintentionally introduced by emitter’s Radio Frequency (RF) front-end components \cite{jagannath_comprehensive_2022, tyler_considerations_2023}. SEI finds broad application in enhancing the security of communication systems and facilitating the detection of illicit user intrusions \cite{shen_deep_2023}.

\begin{figure}[t]
    \centering
    \includegraphics[width=1\columnwidth]{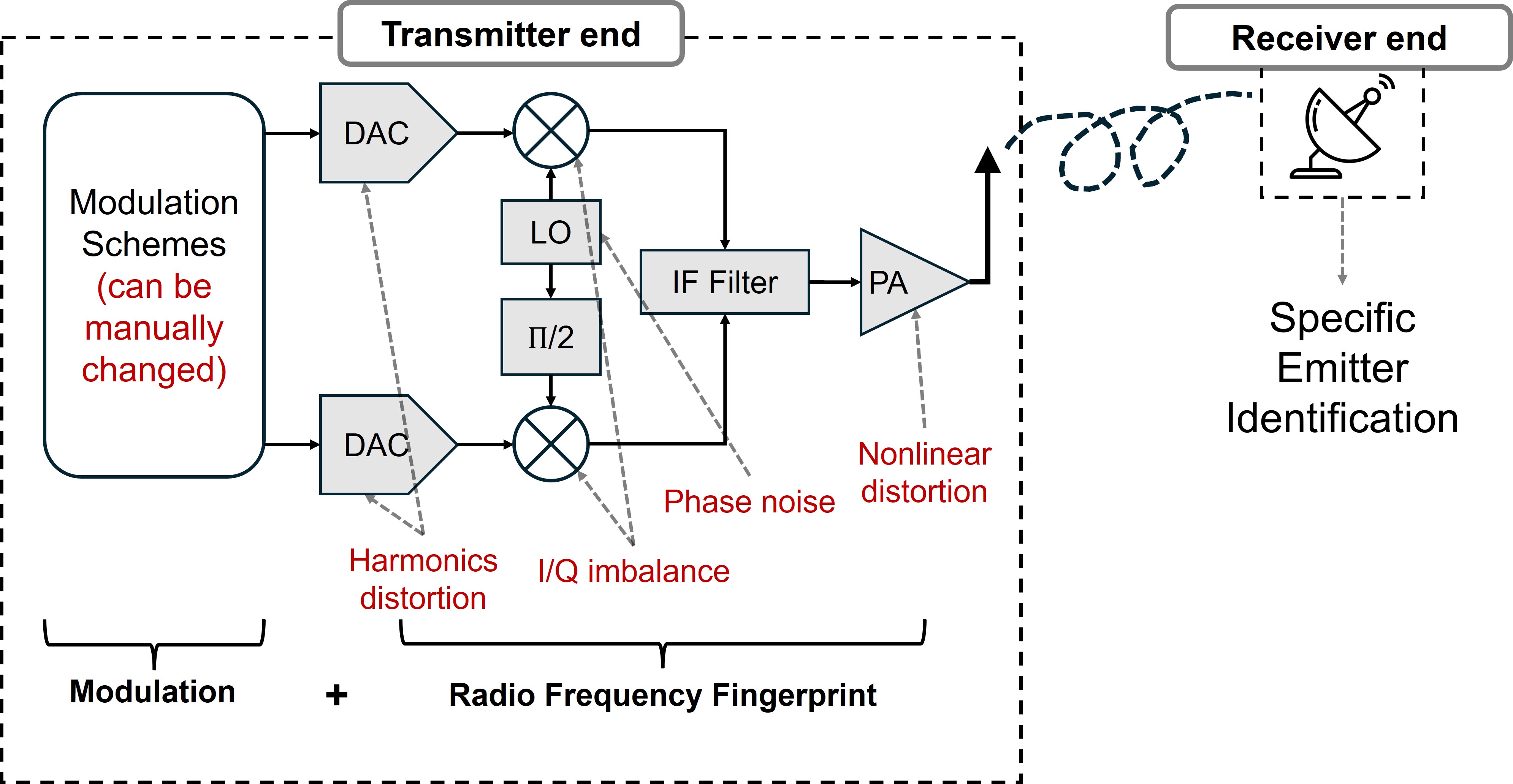}
    \caption{Modulation and the RFF features at the RF front-end.}
    \label{IM_UIM}
    \vspace{-.3cm}
\end{figure}

With the rapid advancement of artificial intelligence, deep learning models have been introduced to SEI, augmenting its ability to capture the RFF features \cite{ma_gafrsnet_2023, yan_intelligent_2023, liao_fast_2024, deng_lightweight_2023}. However, previous research on SEI has predominantly focuses on the signal-level data, which encompasses both modulation features and RFF \cite{tyler_assessing_2022}. As illustrated in Fig. \ref{IM_UIM}, the RF front-end can be delineated into modulation and RFF components. Modulations can be manipulated by humans, involving the modification of amplitude (e.g., Amplitude Modulation), phase (e.g., Phase-Shift Keying), frequency (e.g., Linear Frequency Modulation), and other factors \cite{zang_overview_2018}, while the RFF is primarily shaped by inherent imperfections, rendering it largely immutable. 

The RFF, containing device-specific characteristics, constitutes only a small fraction of the received signal's content. Unlike modulations, the RFF is latent and imperceptible, posing a significant challenge for SEI in isolating it from the signal \cite{liting_unintentional_2022}. Consequently, SEI normally fails in the occurrence of changing modulations. 

It was only recently that scholars began to pay attention to this phenomenon. Tyler, Fadul et al. \cite{tyler_assessing_2022} proposed a method that directly eliminates modulation information from Wi-Fi preamble codes by subtracting an ideal signal, thereby constructing a residual representation. However, this approach requires obtaining the original modulation from the transmitter, which is challenging in practical scenarios. Zhang, Li et al. \cite{zhang_variable-modulation_2022} reconstructed the ideal signal in PSK and QAM modulations, incorporating Unsupervised Domain Adaptation (UDA) methods Domain Adversarial Neural Network (DANN) and Gaussian Encoder (GE) to generate domain-invariant features. Yin, Fu et al. \cite{yin_few-shot_2023} employed Maximum Mean Discrepancy (MMD) to measure the distance between modulation domains. Nevertheless, these works have certain limitations. Firstly, their experimental setups focused solely on PSK or QAM with the same frequency, overlooking the impact of frequency variations. Secondly, reconstructing the ideal signal is highly time-consuming, thereby diminishing the timeliness of SEI. 

\begin{table*}[htbp]
    \centering
    \caption{Brief summary of traditional and modulation-variation SEI methods}
    \setlength{\tabcolsep}{5 mm}
      \begin{tabular}{ccccc}
        \toprule
            & \textbf{Papers} & \textbf{Methods} & \textbf{Ideal Signal} & \textbf{Modulation} \\
        \midrule
      \multirow{4}[0]{*}{Traditional SEI} & Ma 2023 \cite{ma_gafrsnet_2023}     & Resnet   & \multirow{4}[0]{*}{Unrequired} & \multirow{4}[0]{*}{Fixed} \\
            & Yan 2023 \cite{yan_intelligent_2023}     & CVCNN &       &  \\
            & Deng 2023 \cite{deng_lightweight_2023}     & Transformer  &       &  \\
            & Liao 2024 \cite{liao_fast_2024}     & LSTM &       &  \\
        \midrule
      \multirow{4}[0]{*}{Modulation-Variation SEI} & Tyler 2022 \cite{tyler_assessing_2022} & LSTM, CNN & Required & Wi-Fi Preamble \\
            & Zhang 2023 \cite{zhang_variable-modulation_2022} & DANN & Reconstructed & PSK, QAM \\
            & Yin 2023 \cite{yin_few-shot_2023} & MMD   & Unrequired & PSK \\
            & \textbf{Ours}  & MDD   & Unrequired & FM, FSK, PSK, QAM, FSK\_PSK \\
        \bottomrule
      \end{tabular}%
    \label{relatives}%
    \vspace{-.3cm}
  \end{table*}%

In this paper, we explore the benefits of Margin Disparity Discrepancy (MDD) in enhancing SEI in scenarios involving transmitters undergoing the complex and versatile analog and digital modulation schemes.
We establish a framework incorporating MDD through adversarial training, aiming to minimize class distance while simultaneously maximizing domain distance. During classification, we consider samples with varying modulation schemes as the source and target domains, engaging in a minimax game involving MDD and Cross-Entropy (CE) loss. By imposing a rigorous upper bound on domain distance, our approach facilitates the model in discerning domain-invariant RFF features.

Our approach circumvents the necessity of acquiring or constructing ideal signals and demonstrates effectiveness in a broader range of modulation variation scenarios, as depicted in Table \ref{relatives}. The contributions are as follows. 

\begin{itemize}
    \item{We propose MDD to align the varying modulation features, forming minimum intra-class distance and maximum inter-class distance. The feature map visualization demonstrates the effectiveness of our approach.}
    \item{We emit signals with 11 modulation methods through 7 emitters, encompassing both analog and digital modulation schemes, involving variations in amplitude, frequency, and phase, designing a scenario oriented to adaptive modulation variation. The experiments confirm the adaptability of our proposed adversarial SEI framework.}
    \item{The proposed method utilizes the simplest raw signal, while exhibiting superior performance in accuracy and time efficiency compared to other SEI works and domain adaptation methods addressing modulation variation.}
\end{itemize} 

\section{Problem Statement}\label{sec2}

Signal characteristics can be modeled by considering modulations in the ideal signal's \textbf{Amplitude}, \textbf{Frequency} and \textbf{Phase}:
\begin{equation}
\begin{aligned}
r(t) = & \mathcal{M}_{A}(\Delta A, A(t))\exp\{j[2\pi(\mathcal{M}_{f}(f_0, \Delta f, f(t)))t \\ &+ \mathcal{M}_{\phi}(\phi_0, \Delta\phi, \phi(t))]\}+n(t)\text{, }0\leq t\leq T
\end{aligned}
\end{equation}where $\Delta A$ is unintentional amplitude modulation, $f_0$ is the carrier frequency, $\Delta f$ is the unintentionally modulated Carrier Frequency Offset (CFO), $\phi_0$ is the initial phase, $\Delta\phi$ is unintentional phase modulation, $n(t)$ is channel noise, and $T$ is the duration of the signal. $A(t)$, $f(t)$ and $\phi(t)$ are parameters of intentional modulation on amplitude, frequency and phase, while $\mathcal{M}_{A}$, $\mathcal{M}_{f}$ and $\mathcal{M}_{\phi}$ 
represent the implicit factor that cannot be explicitly modeled, leading to the coupling of modulation schemes and the RFF features.

\begin{figure}[!t]
    \centering
    \subfloat[LFM]{\includegraphics[width=.23\columnwidth]{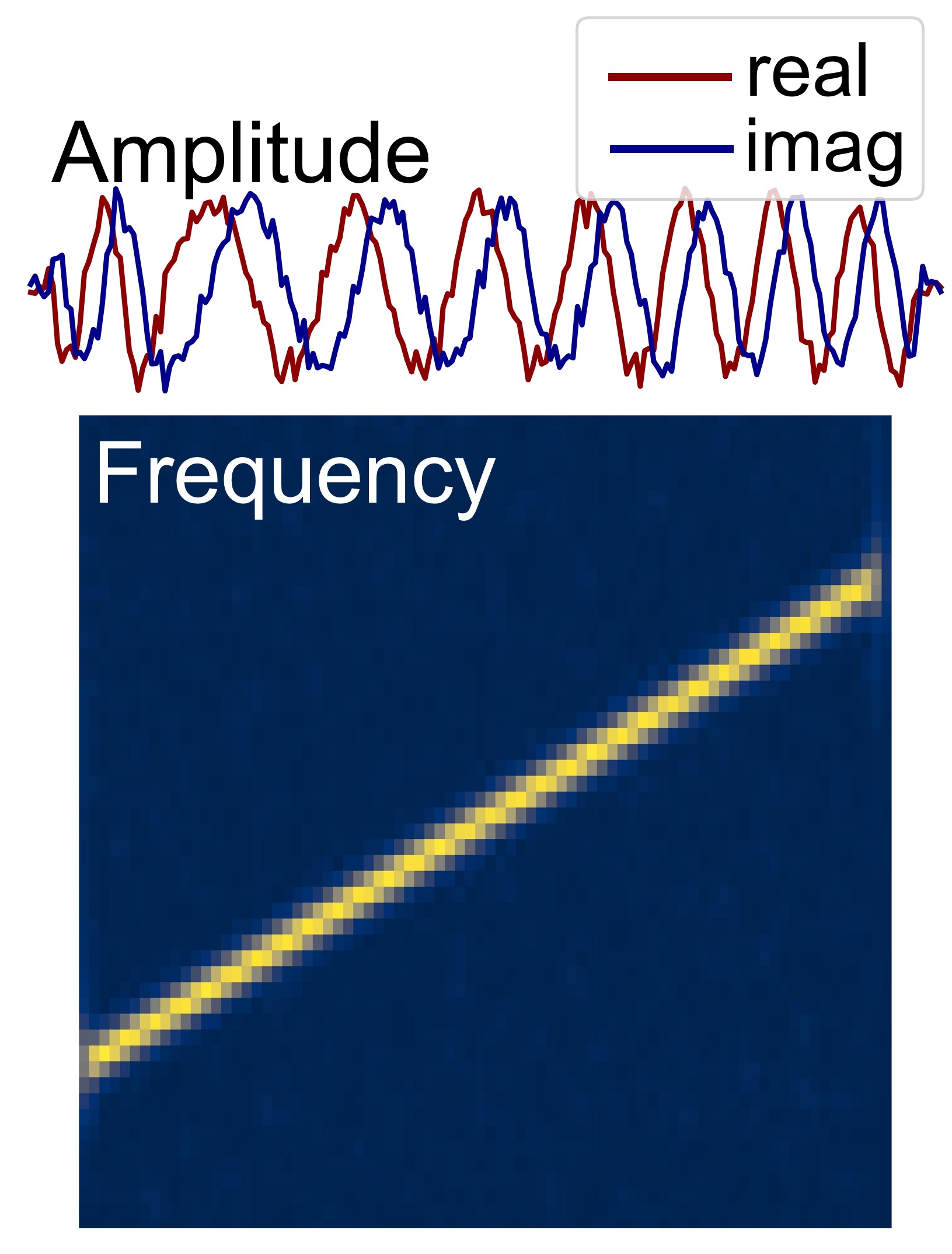}}\hspace{1pt}
    \subfloat[NLFM]{\includegraphics[width=.23\columnwidth]{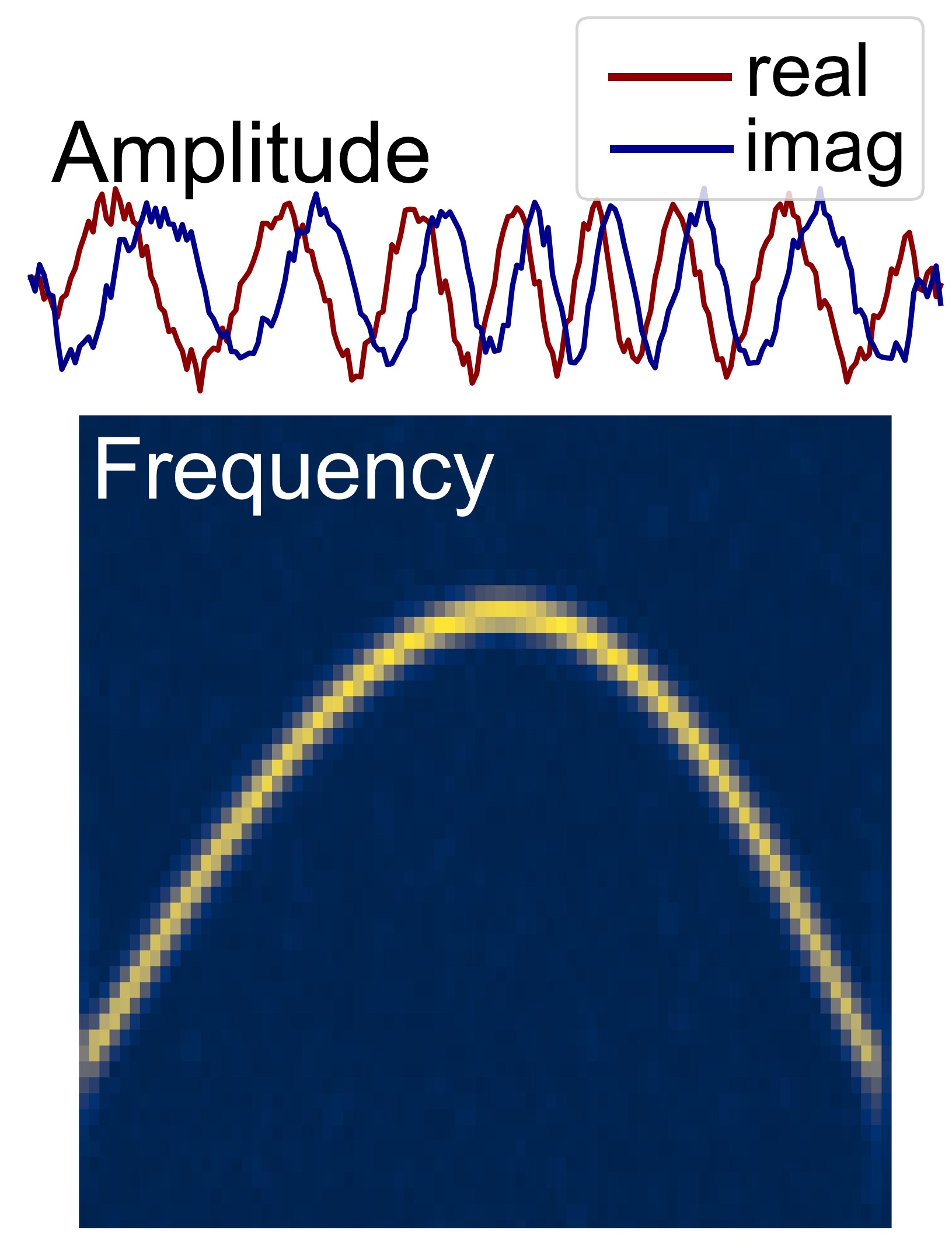}}\hspace{1pt}
    \subfloat[BFSK]{\includegraphics[width=.23\columnwidth]{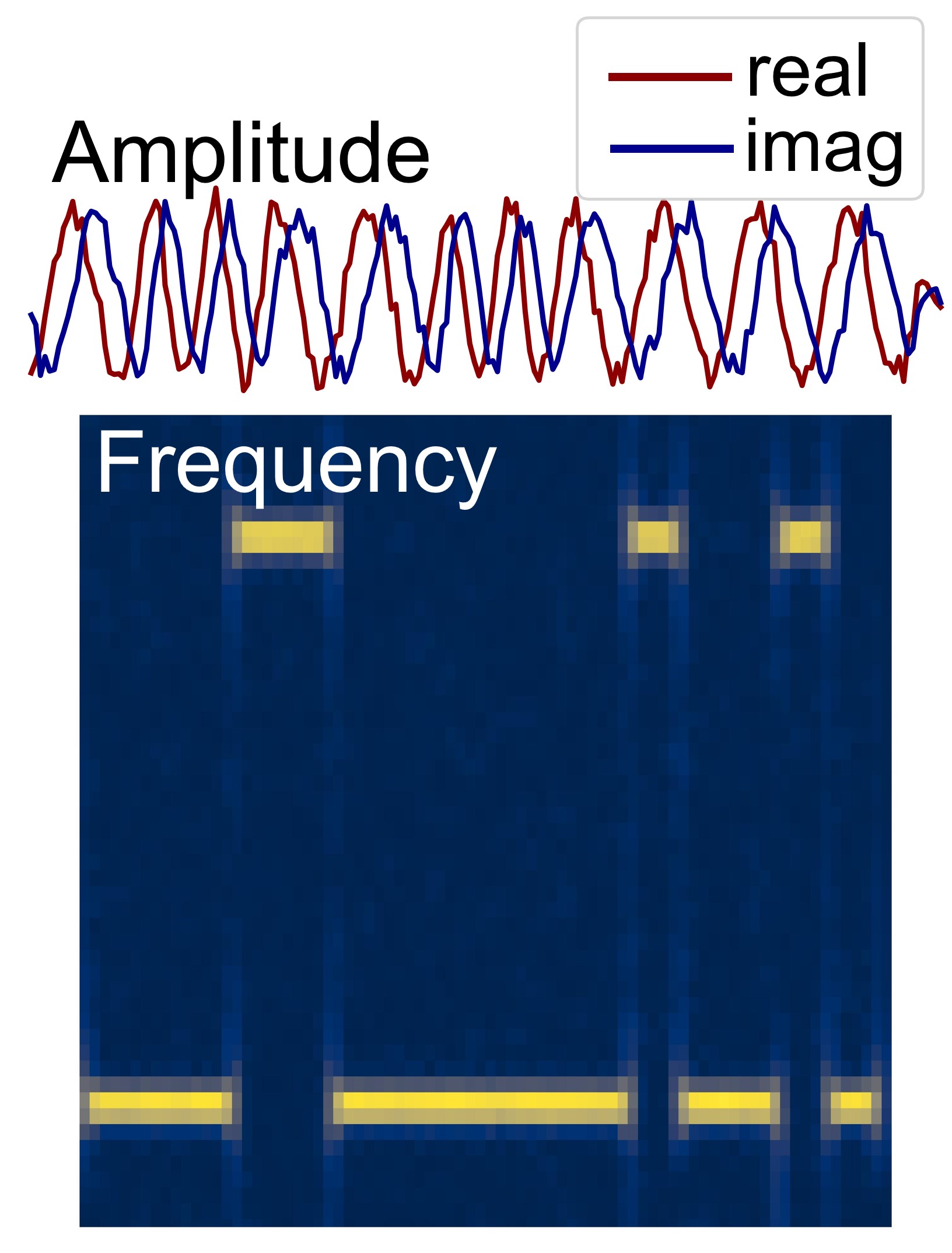}}\hspace{1pt}
    \subfloat[QFSK]{\includegraphics[width=.23\columnwidth]{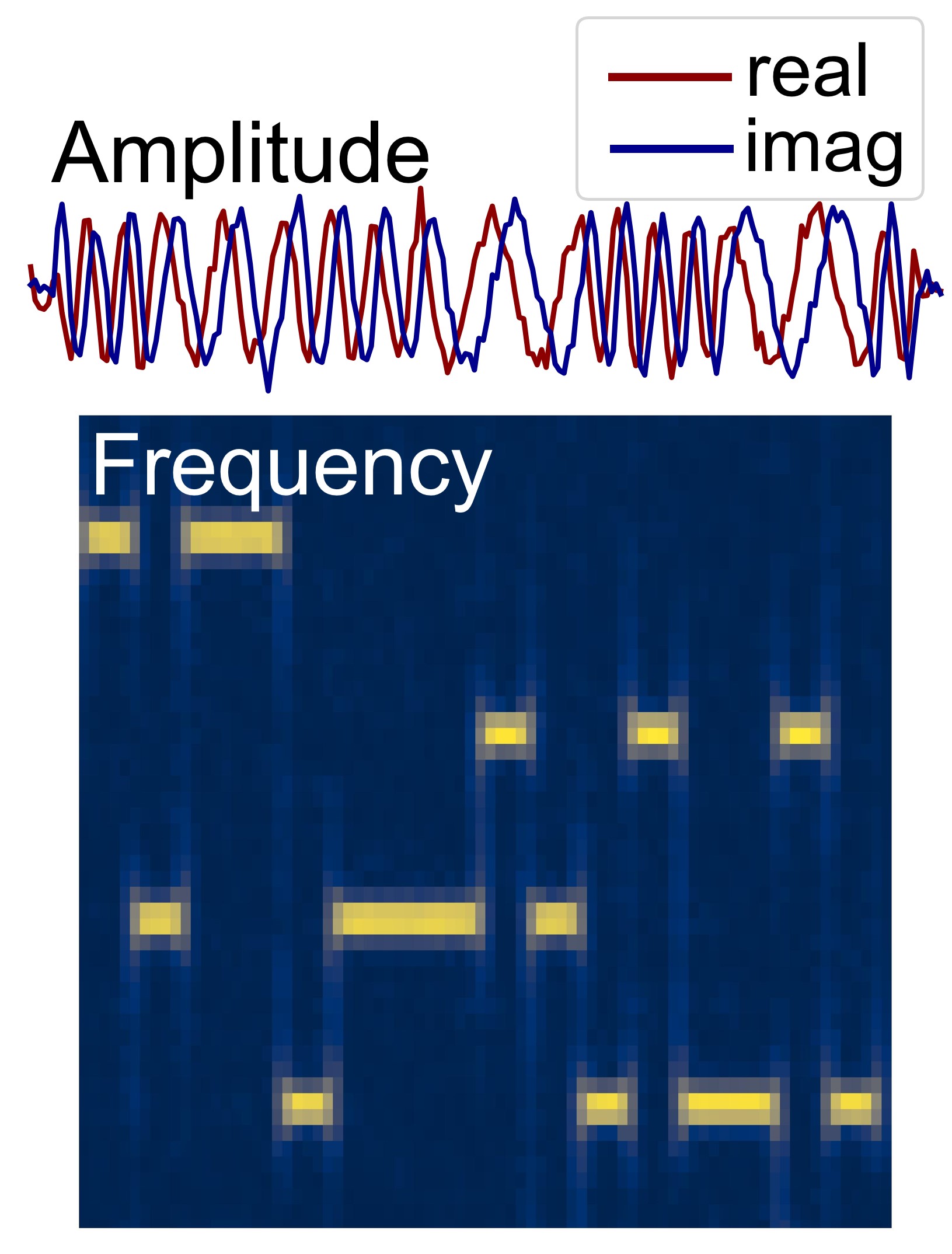}}\\
    \caption{STFT graphs of LFM, NLFM, BFSK and QFSK, reflecting the distinct modulation in frequency features.}\label{stft}
    \vspace{-.3cm}
\end{figure}

\begin{figure}[!t]
    \centering
    \subfloat[BPSK]{\includegraphics[width=.23\columnwidth]{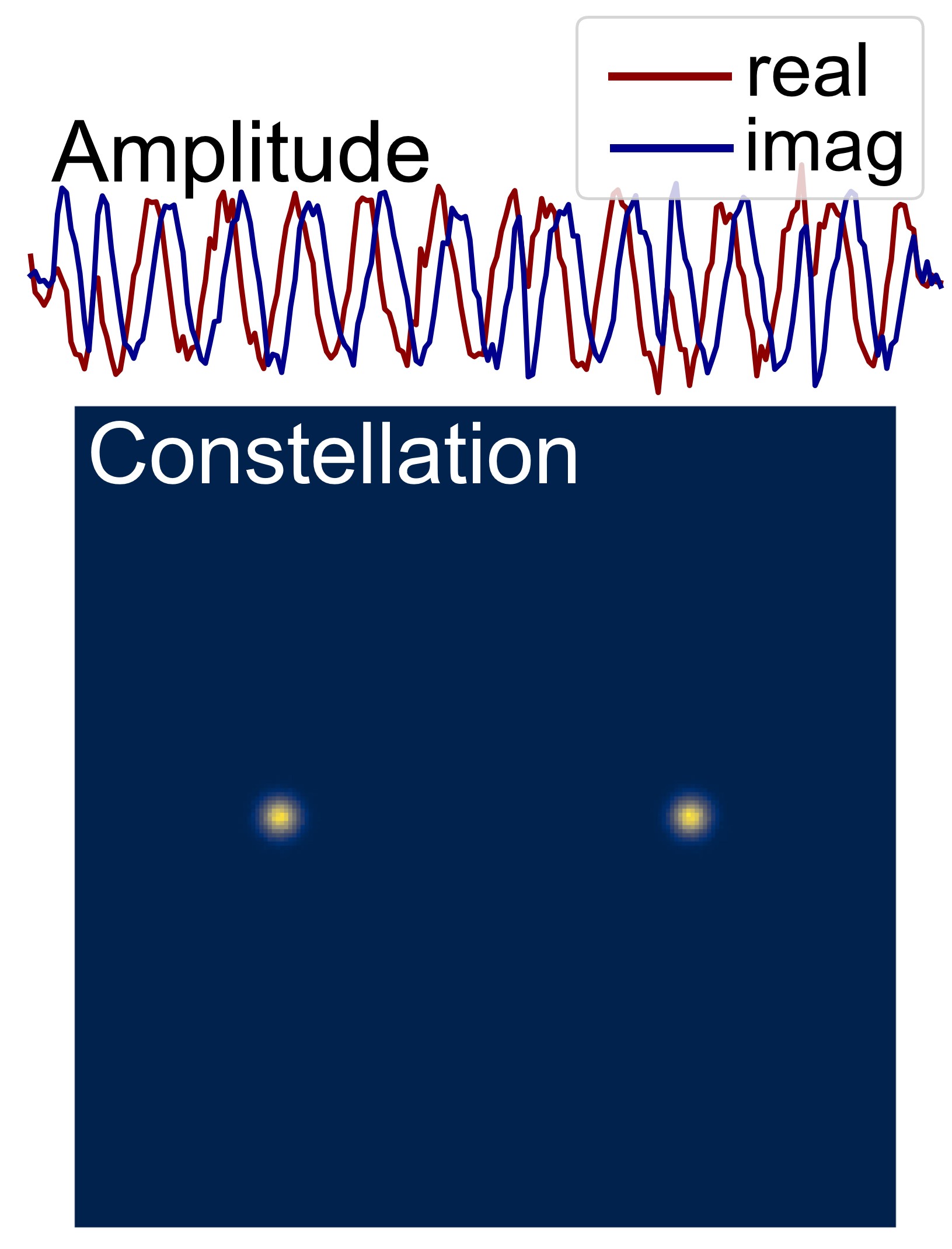}}\hspace{1pt}
    \subfloat[QPSK]{\includegraphics[width=.23\columnwidth]{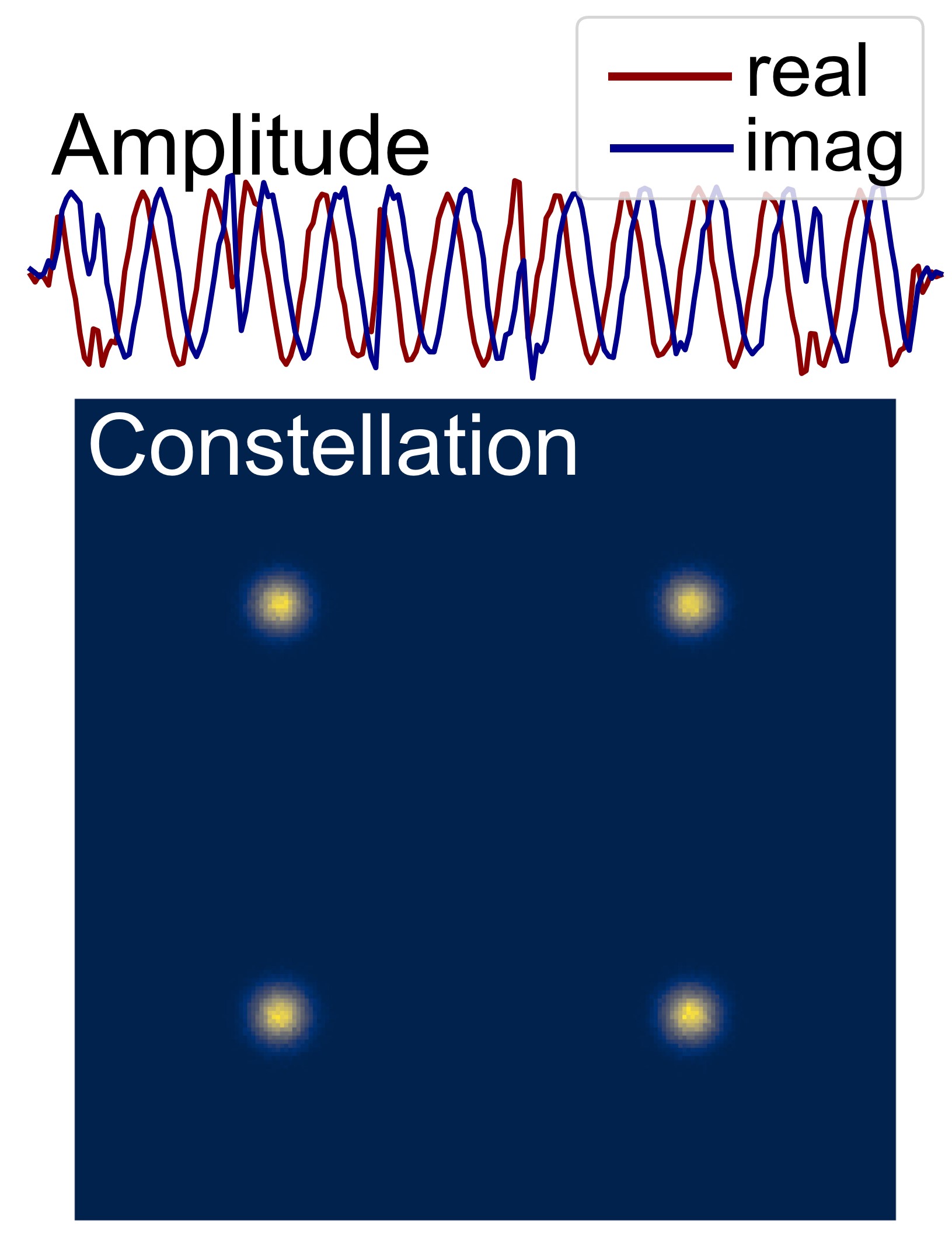}}\hspace{1pt}
    \subfloat[16QAM]{\includegraphics[width=.23\columnwidth]{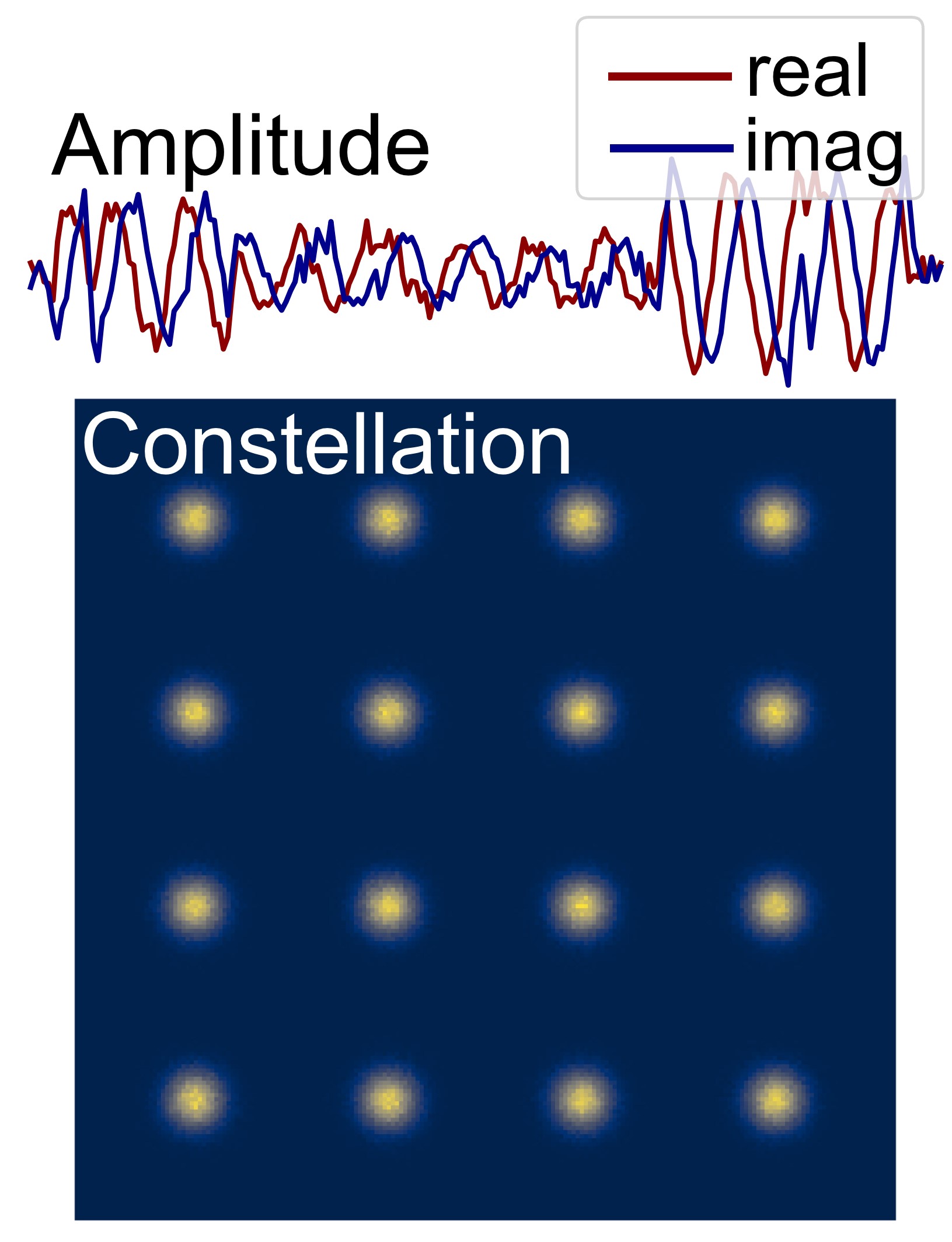}}\hspace{1pt}
    \subfloat[64QAM]{\includegraphics[width=.23\columnwidth]{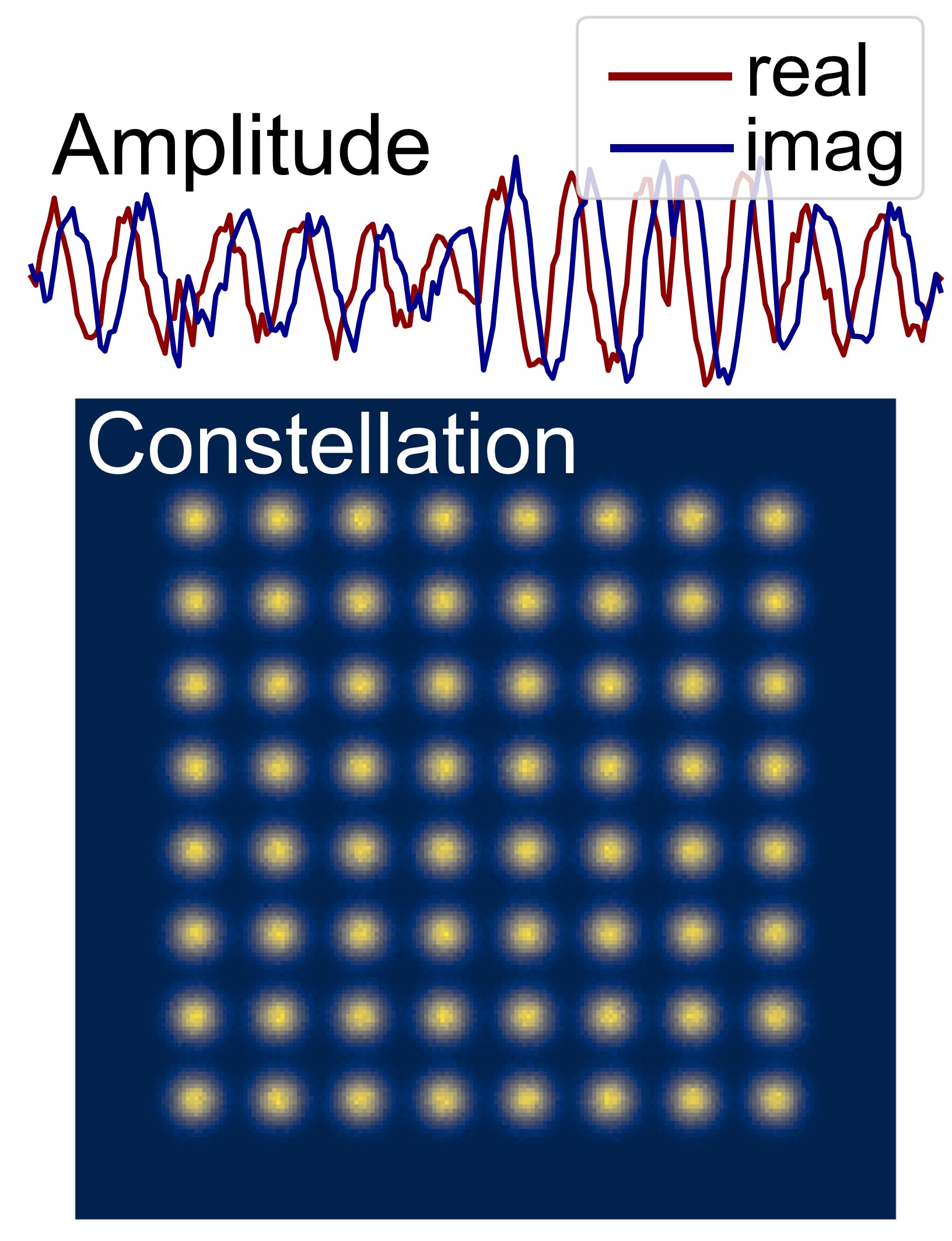}}\\
    \caption{Constellation diagrams of BPSK, QPSK, 16QAM and 64QAM, reflecting the distinct modulation in amplitude and phase features.}\label{constellations}
    \vspace{-.3cm}
\end{figure}

In the ideal SEI scenario, the RFFs such as $\Delta A$, $\Delta f$, and $\Delta \phi$ serves to distinguish transmitters. However, $A(t)$, $f(t)$ and $\phi(t)$ can be modulated more explicit and often disordered in analog and digital modulations. As depicted in Table \ref{checkmarks}, in the case of Frequency Modulations (FM) and Frequency Shift Keyings (FSK), there are variations in $f(t)$; Phase Shift Keyings (PSK) involve variations in $\phi(t)$; In the case of Quadrature Amplitude Modulations (QAM) and Hybrid FSK-PSK, a combination of $A(t)$ and $\phi(t)$, as well as $f(t)$ and $\phi(t)$, can simultaneously occur. The diverse modulations can be discerned on time-frequency domain through Short-Time Fourier Transform (STFT), and on amplitude and phase through constellation diagrams, as shown in Fig. \ref{stft} and Fig. \ref{constellations}. 

\begin{table}[!t]
    \centering
    \fontsize{7.5}{8}\selectfont
    \caption{Variable Features of Different Modulation Schemes}
    \begin{tabular}{llllll}
        \toprule
            & \textbf{FM} & \textbf{FSK}  & \textbf{PSK}  & \textbf{QAM}  & \textbf{FSK\_PSK} \\
        \midrule
    Amplitude &       &       &       & \checkmark     &  \\
    Frequency &\checkmark & \checkmark     &     &       & \checkmark \\
    Phase &       &       &\checkmark        & \checkmark     & \checkmark \\
    Discrete symbols&       & \checkmark     & \checkmark     & \checkmark     & \checkmark \\
    \bottomrule
    \end{tabular}%
    \label{checkmarks}%
    \vspace{-.3cm}
\end{table}%

\begin{figure*}[t]
    \centering
    \includegraphics[width=2\columnwidth]{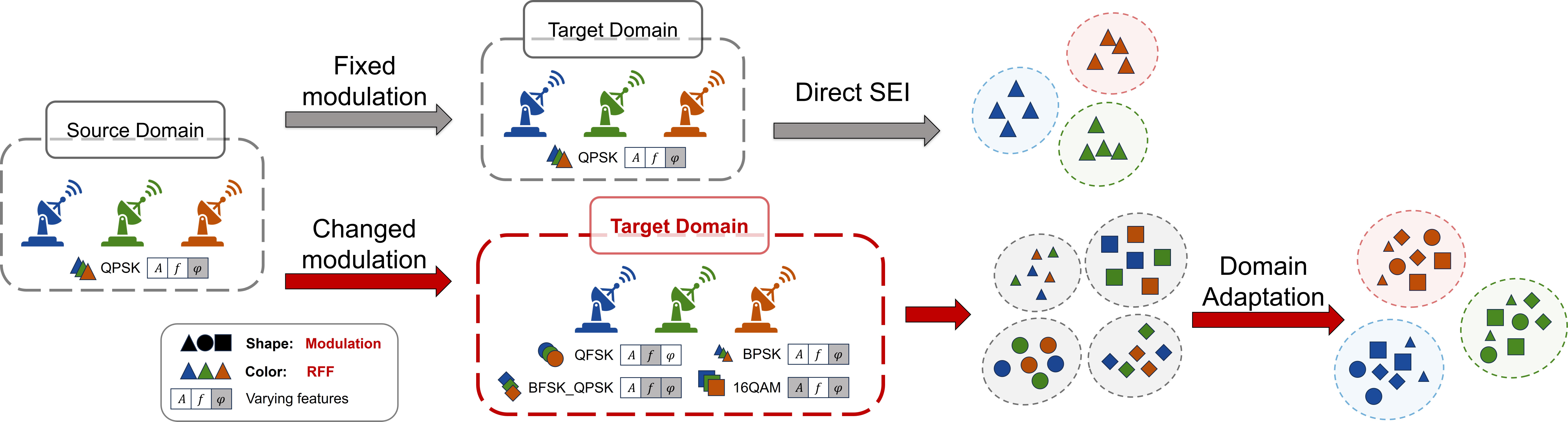}
    \caption{SEI with fixed and variable modulation schemes.}
    \label{problem}
    \vspace{-.3cm}
\end{figure*}

When both training and testing samples belong to modulation schemes depicted in either Fig. \ref{stft} or Fig. \ref{constellations}, SEI is often achievable. However, there are scenarios that modulation scheme undergoes significant change. For example, labeled samples are derived from QAM (varying amplitude and phase), while unlabeled samples originate from FSK (varying frequency), as depicted in Fig. \ref{problem}.


\begin{figure}[!t]
    \centering
    \subfloat[]{\includegraphics[width=.315\columnwidth]{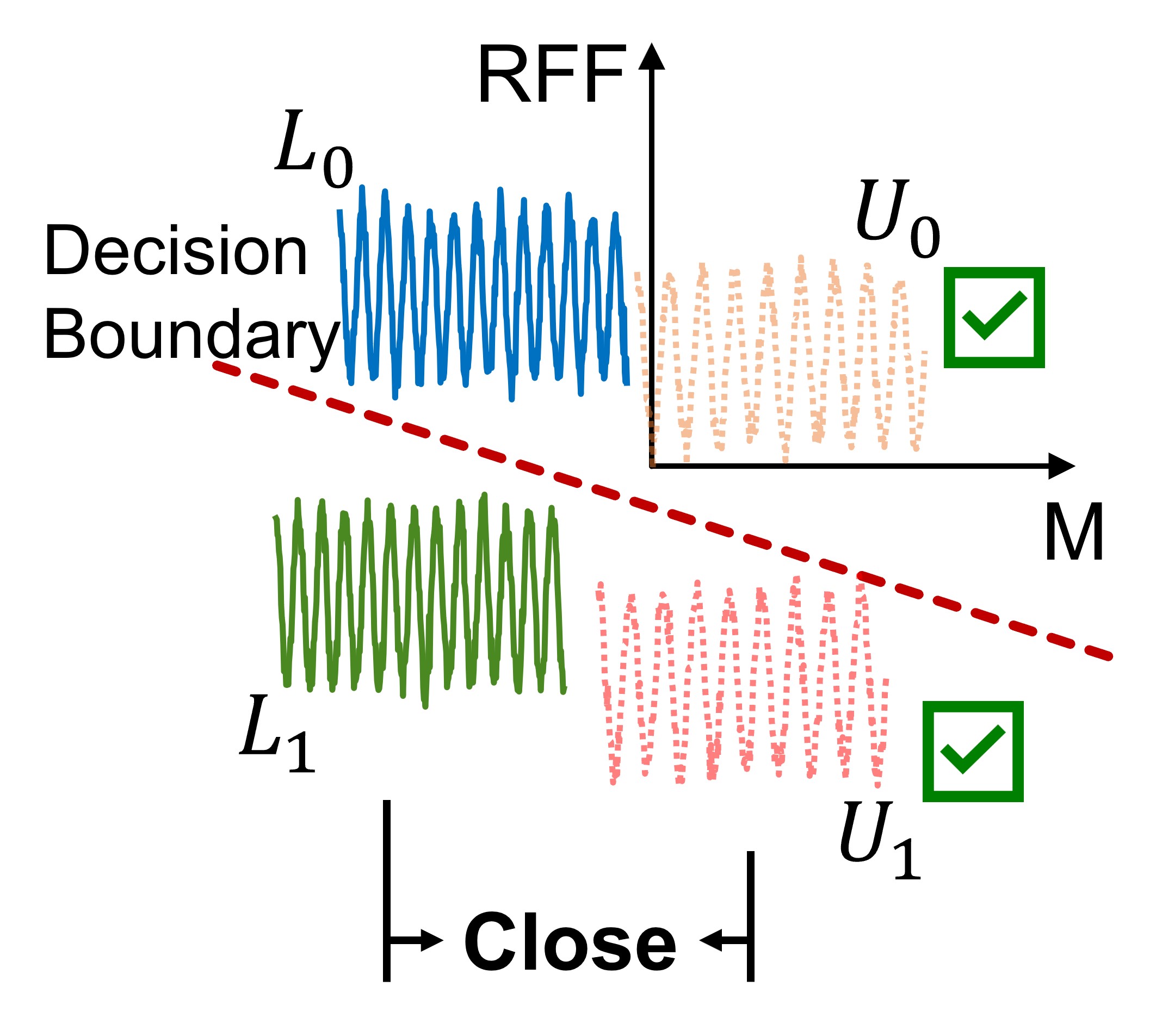}\label{da}}
    \subfloat[]{\includegraphics[width=.372\columnwidth]{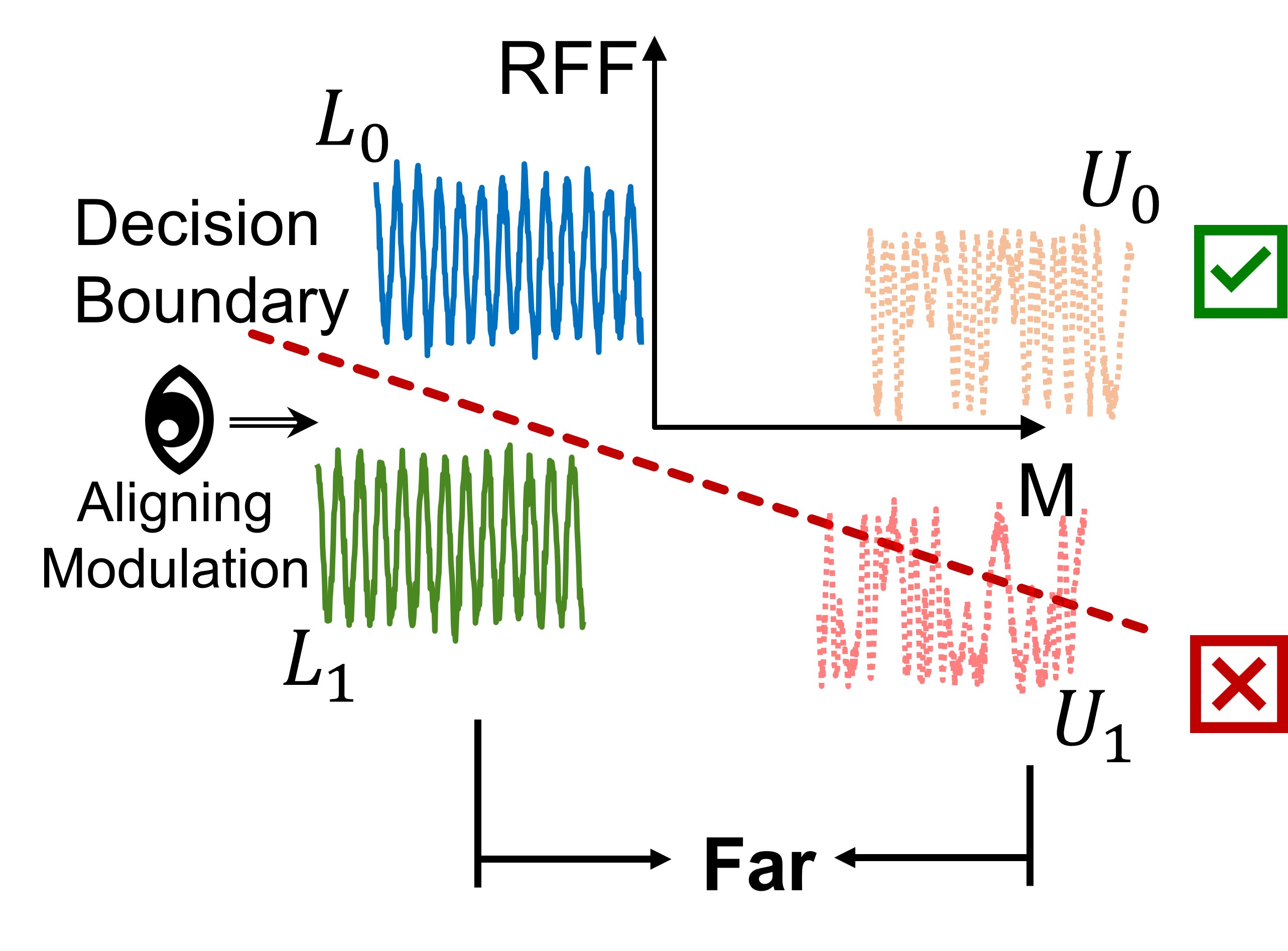}\label{db}}
    \subfloat[]{\includegraphics[width=.293\columnwidth]{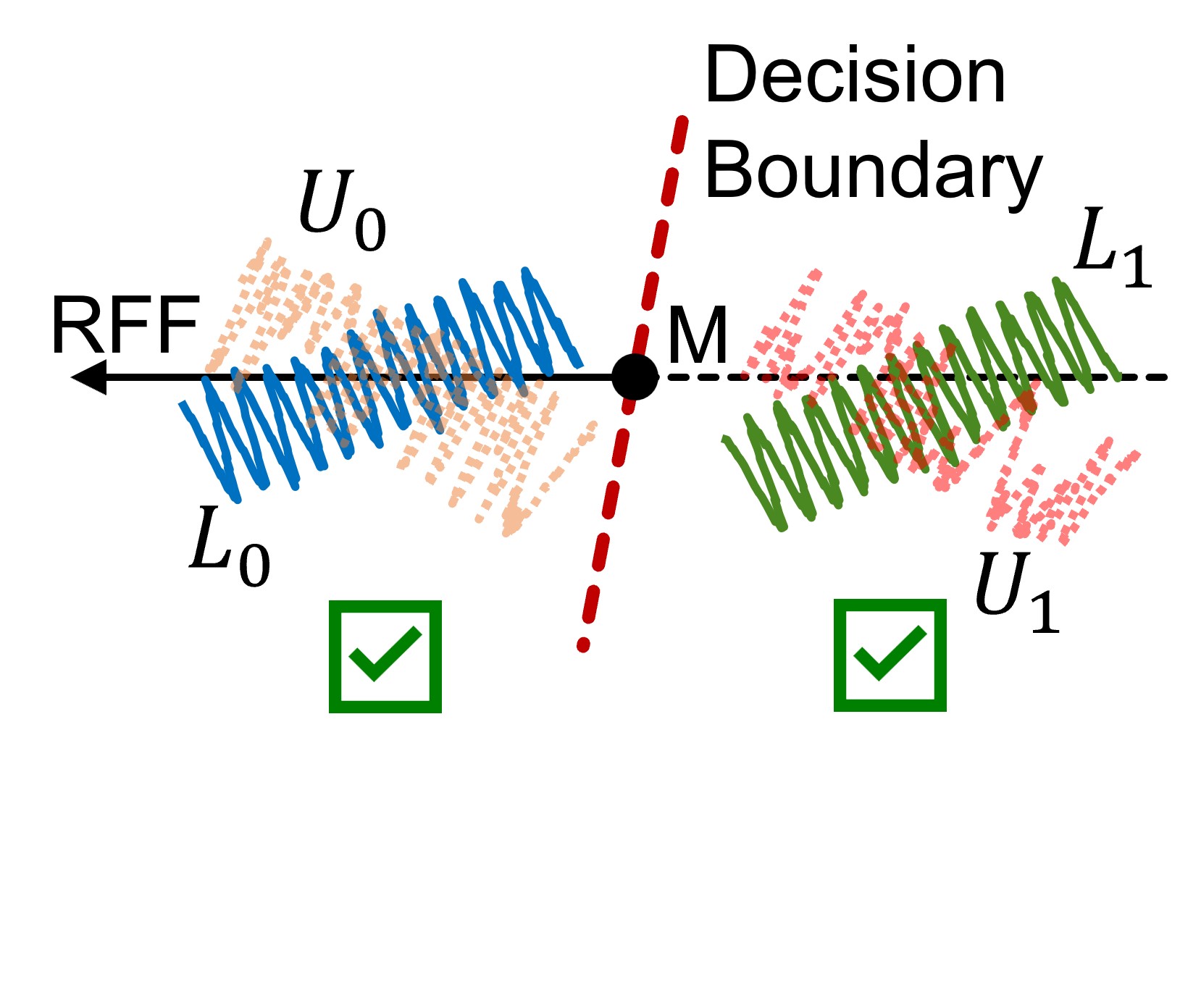}\label{dc}}
    \caption{SEI decision boundary influenced by modulation variation (M denotes modulation).}\label{decisions}
    \vspace{-.3cm}
\end{figure}

In the face of varying degrees of changes in modulation schemes, a classifier will delineate decision boundaries with pronounced distinctiveness. Fig. \ref{decisions} illustrates this challenge with a binary classification scenario. Here, $(L_0, U_0)$ and $(L_1, U_1)$ denote two emitters labeled as `0' and `1'. $L$ and $U$ refers to Labeled and Unlabeled samples. Establishing coordinate axes for modulation and the RFF, we assume that $L_0$ shares a set of RFFs with $U_0$, and likewise for $L_1$ and $U_1$, indicating their origin from two distinct transmitters. However, $L_0$ and $L_1$ are highly similar in modulation, as do  $U_0$ and $U_1$. 

As depicted in Fig. \ref{da}, when the modulation features between $(L_0, L_1)$ and $(U_0, U_1)$ are close, the impact of such decision boundary is minimal, enabling correct classification of $U_0$ and $U_1$. However, in the scenario illustrated in Fig. \ref{db}, a significant disparity exists in modulation features between labeled and unlabeled samples (e.g., QAM and FSK), leading to misclassification of $U_1$. Our objective is to develop a classifier capable of aligning modulation features by focusing on the "eye" in Fig. \ref{db}, which gazes towards the modulation axis. From this perspective, the feature space transforms from Fig. \ref{db} to Fig. \ref{dc}. In this transformed scenario, the modulation axis undergoes compression, thereby enabling the model to establish a more effective decision boundary.

\section{SEI Based on Margin Disparity Discrepancy}\label{sec3}

We designate $N$ labeled samples with one modulation as source domain ${\mathcal{S}} = \{(x^{s}_i, y^{s}_i)\}^{N}_{i=1}$ and $M$ unlabeled samples with another modulation as target domain ${\mathcal{T}} = \{(x^{t}_j)\}^{M}_{j=1}$. In the hypothesis space $\mathcal{H}$, our aim is to find a hypothesis $h:\mathcal{X}\mapsto\mathcal{Y}$ with minimum error in target domain $\epsilon_{\mathcal{T}}(h)=\mathbb{E}_{(\mathbf{x}^t,\mathbf{y}^t)\sim\mathcal{T}}[\mathcal{L}(h(\mathbf{x}^t),\mathbf{y}^t)]$, where $\mathcal{L}$ is a loss function. We also define the error in source domain as $\epsilon_{\mathcal{S}}(h)=\mathbb{E}_{(\mathbf{x}^t,\mathbf{y}^t)\sim\mathcal{S}}[\mathcal{L}(h(\mathbf{x}^t),\mathbf{y}^t)]$. Due to the absence of labels in the target domain, we can only approximate the limitation of $\epsilon_{\mathcal{T}}$ by utilizing $\epsilon_{\mathcal{S}}$, along with assessing the distance between the source and target domains.

In general, it is commonly assumed that $\mathcal{L}$ is symmetric and satisfies the triangle inequality. The discrepancy between two hypotheses $h$ and $h'$ can be expressed as
\begin{equation}\begin{aligned}\epsilon_{\mathcal{D}}(h,h^{\prime})=\mathbb{E}_{(\mathbf{x},\mathbf{y})\sim\mathcal{D}}[\mathcal{L}(h(\mathbf{x}),h^{\prime}(\mathbf{x}))]\end{aligned}\label{err_d}\end{equation}

In domain adaptation, a common assumption known as the \textit{impossibility theorem} posits that achieving domain adaptation is not possible unless there exists a sufficiently small \textit{ideal joint error}, represented as the minimum sum of the source risk and target risk \cite{david_impossibility_2010}
\begin{equation}
    \begin{aligned}
        h^{*} &\triangleq \arg\min_{h\in\mathcal{H}}\left[\epsilon_{\mathcal{S}}\left(h\right)+\epsilon_{\mathcal{T}}\left(h\right)\right]\\
        \epsilon_{ideal} &= \epsilon_{\mathcal{S}}\left(h^*\right)+\epsilon_{\mathcal{T}}\left(h^*\right)
    \end{aligned}
\end{equation}where $h^{*}$ represents the hypothesis in $\mathcal{H}$ that minimizes the joint error.  

Therefore, by utilizing Equation \ref{err_d}, we can obtain an upper bound representation for the target risk
\begin{equation}
    \begin{aligned}
        d_{h,\mathcal{H}}(\mathcal{S},\mathcal{T}) &\triangleq \left|\epsilon_{\mathcal{S}}\left(h,h^*\right)-\epsilon_{\mathcal{T}}\left(h,h^*\right)\right| \\
        \epsilon_{\mathcal{T}}
        \left(h\right) & \leq \epsilon_{\mathcal{S}}\left(h\right)
        +\left[\epsilon_{\mathcal{S}}\left(h^{*}\right)
        +\epsilon_{\mathcal{T}}\left(h^{*}\right)\right] + d_{h,\mathcal{H}}(\mathcal{S},\mathcal{T})
    \end{aligned}
\end{equation}where $\left|\epsilon_{\mathcal{S}}\left(h,h^*\right)-\epsilon_{\mathcal{T}}\left(h,h^*\right)\right|$ is defined the Disparity between ${\mathcal{S}}$ and ${\mathcal{T}}$ as $d_{h,\mathcal{H}}(\mathcal{S},\mathcal{T})$.

In order to minimize $\epsilon_{\mathcal{T}}$, our objective is to reduce the $d_{h,\mathcal{H}}(\mathcal{S},\mathcal{T})$. However, due to the lack of labels in $\mathcal{T}$, we cannot ascertain the value of $h^*$ and  precisely estimate the $d_{h,\mathcal{H}}(\mathcal{S},\mathcal{T})$.

\begin{figure*}[t]
    \centering
    \includegraphics[width=1.95\columnwidth]{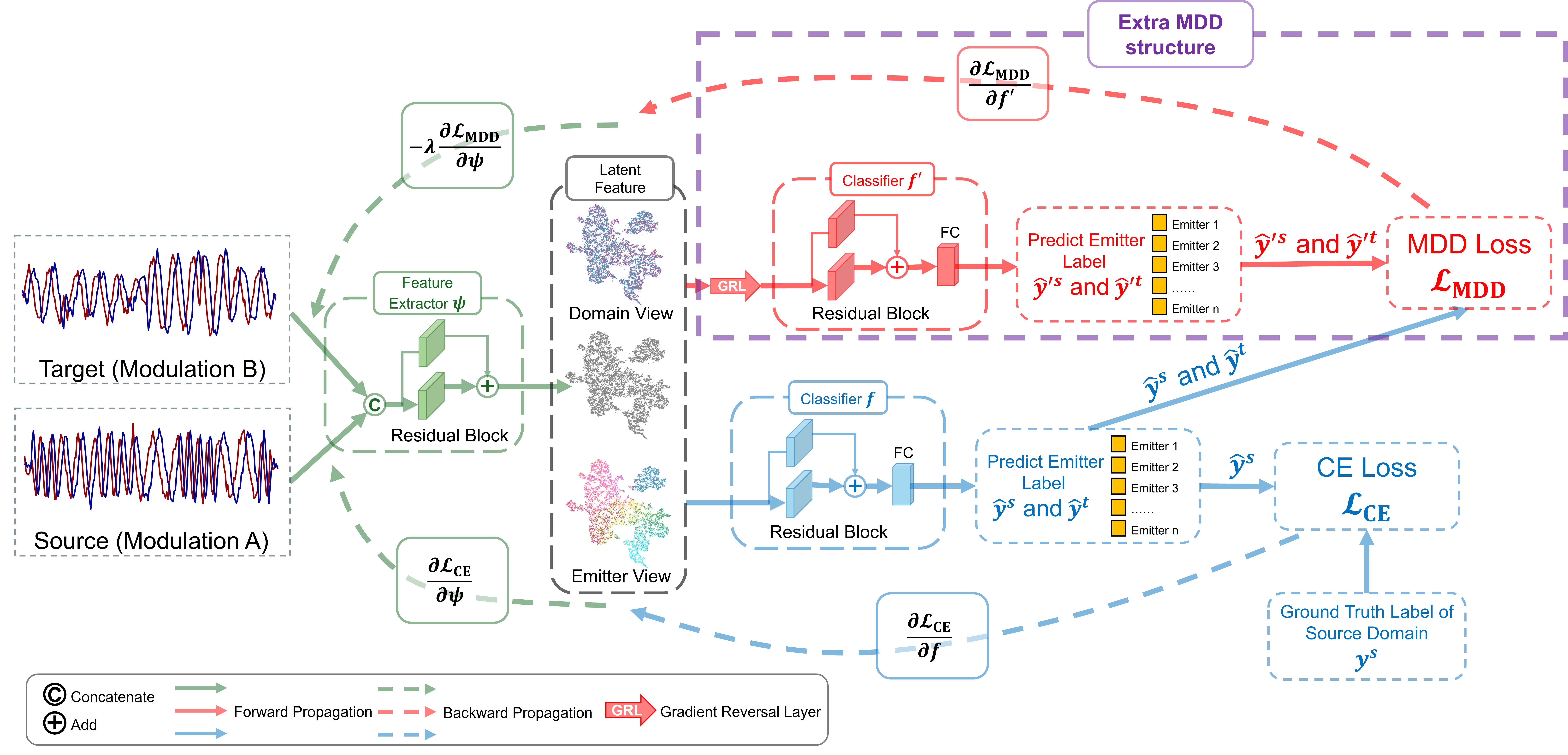}
    \caption{The structure for MDD algorithm implementation. Emitter 1$\sim$n represent the identified transmitter labels. During model training, test data of target modulation (Modulation B) are utilized without labels. The classifier simultaneously outputs predicted values for both source and target domain samples.}
    \label{MDD_structure}
    \vspace{-.3cm}
\end{figure*}

The Margin Disparity Discrepancy (MDD) \cite{zhang_MDD_2019} provides a rigorous boundary, significantly simplifies the $d_{h,\mathcal{H}}(\mathcal{S},\mathcal{T})$. MDD introduces $f'$ as an adversarial classifier, which shares the same hypothesis space  $\mathcal{F}$ with classifier $f$. Denote the margin of $f$ at one labeled sample $(x, y)$ as
\begin{equation}
    \begin{aligned}
        \rho_f(x,y)\triangleq\frac{1}{2}(f(x,y)-\max_{y^{\prime}\neq y}f(x,y^{\prime}))
    \end{aligned}
\end{equation}
and the margin loss as
\begin{equation}
    \begin{aligned}
        \Phi_\rho(x)\triangleq\begin{cases}0&\rho\le x\\1-x/\rho&0\le x\le\rho\\1&x\le0\end{cases}
    \end{aligned}
\end{equation}
then Margin Disparity is defined distinct from Equation \eqref{err_d} with the margin $\rho$ as
\begin{equation}
    \begin{aligned}
        \epsilon_{\mathcal{D}}^{(\rho)}(f^{\prime},f)=\mathbb{E}_{(x,y)\thicksim\mathcal{D}}[\Phi_\rho(\rho_{f^{\prime}}(x,h_f(x))]
    \end{aligned}
\end{equation}where $ h_{f}: x \mapsto\text{arg}\max_{y\in\mathcal{Y}}f(x,y)$ denotes the labeling function induced by $f$.

With the definition $\gamma\triangleq e^ {\rho}$, the upper bound of $d_{f, \mathcal{F}}(\mathcal{S}, \mathcal{T})$ is estimated with the following equation defined as MDD \cite{zhang_MDD_2019}. 
\begin{equation}
    \begin{aligned}
        \mathcal{L}_{\mathrm{MDD}}(f,\psi) &=\mathbb{E}_{\mathbf{x}^{s}\sim{\mathcal{S}}}\left[\mathcal{L}(f'(\psi(\mathbf{x}^{s})), f(\psi(\mathbf{x}^{s})))\right]\\& - \gamma \mathbb{E}_{\mathbf{x}^{t}\sim{\mathcal{T}}}\left[\mathcal{L}'(f'(\psi(\mathbf{x}^{t})), f(\psi(\mathbf{x}^{t})))\right]
    \end{aligned}\label{L_MDD}
\end{equation}
where $\psi$ denotes a feature extractor. $\gamma$ constrains $f'$ within a reasonable range, preventing the overestimation of the generalization bound.

In the source domain, we utilize CE Loss as the loss function. In the target domain, we employ modified CE loss to mitigate potential issues such as gradient explosion or vanishing that may arise during adversarial learning.
\begin{equation}
    \begin{aligned}
        \mathcal{L}(f'(\psi(x^s)),f(\psi(x^s)))\triangleq-\log[\sigma_{h_f(\psi(x^s))}(f'(\psi(x^s)))]\\
    \end{aligned}\label{L_CE}
\end{equation}
\begin{equation}
    \begin{aligned}
        \mathcal{L}'(f'(\psi(x^t)),f(\psi(x^t)))\triangleq\log[1-\sigma_{h_f(\psi(x^t))}(f'(\psi(x^t)))]
    \end{aligned}
\end{equation}
where $\sigma$ is the \textit{softmax} function
\begin{equation}
    \begin{aligned}
        \sigma_j(\mathbf{z})=\frac{e^{z_j}}{\sum_{i=1}^ke^{z_i}},\mathrm{~for~}j=1,\ldots,k,{ }\mathbf{z}\in\mathbb{R}^k
    \end{aligned}
\end{equation}

We utilize $f'$ to make correct predictions in the source domain (as reflected in the error $\epsilon_{\mathcal{S}}$), while ensuring that it provides different predictions in the target domain compared to $f$ (as reflected in the upper bound of $d_{f,\mathcal{F}}(\mathcal{S},\mathcal{T})$). This \textit{minimax game} can be formulated as
\begin{equation}
\begin{aligned}
    \begin{gathered}\min_{f,\psi}\epsilon_{\mathcal{S}}+\eta \mathcal{L}_{\mathrm{MDD}}\\
    \max_{{f'}}\mathcal{L}_{\mathrm{MDD}}\end{gathered}
\end{aligned}
\end{equation}where $\eta$ is the trade-off coefficient between the error in source domain and MDD loss function.

\begin{algorithm}[!t]
    \caption{Proposed method}\label{algo1}
    {TRAINGNING PROCEDURE: }
    \begin{algorithmic}[1]
        
        \Require{labeled samples with certain modulations ${\mathcal{S}} = \{(x^{s}_i, y^{s}_i)\}^{N}_{i=1}$ and unlabeled samples with different modulations ${\mathcal{T}} = \{(x^{t}_i, )\}^{M}_{i=1}$.}

        \State {Samples from ${\mathcal{S}}$ and ${\mathcal{T}}$ are fed to \textbf{Feature Extractor} $\psi$ to obtain $\psi(x^{s}_i)$ and $\psi(x^{t}_i)$.} \label{algo_feature}
    
        \State {\textbf{Classifier} $f$ and \textbf{Adversarial Classifier} $f'$ are employed to form 
            \begin{itemize}
                \item $\widehat{y}^{{s}} = f(\psi(x^{s}_i))$, $\widehat{y}^{{t}} = f(\psi(x^{t}_i))$
                \item $\widehat{y}'^{{s}} = f'(\psi(x^{s}_i))$, $\widehat{y}'^{{t}} = f'(\psi(x^{t}_i))$
            \end{itemize}\label{algo_y}}
        
        \State {Calculate $\mathcal{L}$ and $\mathcal{L}'$ with $\widehat{y}^{s}$ and $y^{s}$ using Equation \eqref{L_CE}. Calculate $\mathcal{L}_\mathrm{MDD}$ with $\widehat{y}^{{s}}$, $\widehat{y}^{{t}}$, $\widehat{y}'^{{s}}$ and $\widehat{y}'^{{t}}$ using Equation \eqref{L_MDD}.}
    
        \State {Compute the complete loss function utilizing Equation \eqref{complete_L}, and train the whole network until convergence.}
        \Statex {}
    \end{algorithmic}

    {PREDICTING PROCEDURE: }
\begin{algorithmic}[1]
    
    \Require Unlabeled samples ${\mathcal{T}} = \{(x^{t}_i, )\}^{M}_{i=1}$. 
    \State {Follow step 1 and 2} in the training procedure.
    \Ensure $\widehat{y}^{{t}}$ is the prediction.
\end{algorithmic}
\end{algorithm}

Due to the challenge of optimizing the margin loss through gradient descent, the objectives of minimizing and maximizing are combined into a unified form with a trade-off parameter $\lambda$. Minimizing the following term reduces the target error when the adversarial classifier $f'$ approaches the upper bound.
\begin{equation}
    \begin{aligned}
    \min_{\psi,f}\mathbb{E}_{(\mathbf{x}^s,\mathbf{y}^s)\sim{\mathcal{S}}}\mathcal{L}(f(\psi(\mathbf{x}^s)),\mathbf{y}^s)+\lambda \mathcal{L}_{\mathrm{MDD}}(f,\psi)
    \end{aligned}\label{complete_L}
\end{equation}

The overall framework of the method is illustrated in Fig. \ref{MDD_structure}. As the loss is non-differentiable in the classifier part, we employ Gradient Reversal Layer (GRL) \cite{bach_unsupervised_2015} to train the feature extractor in order to minimize the disparity loss term. The implementation of MDD on cross-modulation SEI is summarized in Algorithm \ref{algo1}. As training progresses, $\psi ({\mathcal{T}})$ will gradually approach $\psi ({\mathcal{S}})$, aligning domain-variant features.

\section{Experiment}\label{sec4}

\subsection{Data Collection}\label{subsec5_1}

\begin{table}[htbp]
    \centering
    \caption{Details of Data Collection}
    \begin{threeparttable}
        \begin{tabular}{cc}
    \toprule
      \textbf{Items} & \textbf{Setting} \\
    \midrule
      Format & I/Q Signal \\
      Transmitter & 7 HackRF One \\
      Receiver & 1 USRP B210 \\
      Radio frequency & 330 MHz \\
      Carrier frequency (Digital) & 1 MHz \\
      Carrier frequency (Analog) & 0.8 MHz to 1.2 MHz \\
      Signal-to-Noise Ratio & 15 dB \\
      Pulse width  & 12 $\mu$s \\
      Pulse detect algorithm & Self autocorrelation \\
      Sampling rate & 16 MHz \\
      Symbol rate (Digital) & 1 MHz \\
      Sample length & 200 \\
    \bottomrule
      \end{tabular}%
    \end{threeparttable}
    \label{data_collection}%
    \vspace{-.3cm}
  \end{table}%

\begin{table}[htbp]
    \centering
\begin{threeparttable}
    \caption{Experiment Modulations}
    \begin{tabular}{llllll}
\toprule
    Modulation & FM   & FSK  & PSK  & QAM  & FSK\_PSK \\
\midrule
    Sub Schemes & \makecell{LFM\\NLFM} & \makecell{BFSK\\QFSK} & \makecell{BPSK\\QPSK\\8PSK} & \makecell{16QAM\\64QAM} & \makecell{BFSK\_QPSK\\QFSK\_BPSK} \\
\bottomrule
    \end{tabular}
    \label{experiment_modulation}
\end{threeparttable}

\end{table}

We divided the experiment sets with modulation variation into 5 groups: FM, FSK, PSK, QAM, and Hybrid FSK-PSK. Each group comprises 1 labeled modulation type for supervised training (Source) and 4 unlabeled modulation types (Target) for unsupervised learning and testing. 

\subsection{Implementation Details}\label{subsec5_2}

\begin{table}[htbp]
    \centering
    \caption{Details of Parameters}
    \begin{threeparttable}
      \begin{tabular}{ccc}
    \toprule
      \textbf{Parameters} & \textbf{Source Domain} & \textbf{Target Domain} \\
    \midrule
      Number of Modulations & 2 or 3   & 1 \\
      No. training samples & 4,000 (labeled) & 2,000 (unlabeled) \\
      No. testing samples & None  & 2,000 (unlabeled) \\
    \midrule
      Dimension of Samples & \multicolumn{2}{c}{I/Q signal (2, 200)} \\
      Number of Categories & \multicolumn{2}{c}{7} \\
      Batch size & \multicolumn{2}{c}{32} \\
      Epochs (Pretrain\tnote{1} ) & \multicolumn{2}{c}{30} \\
      Epochs  & \multicolumn{2}{c}{100} \\
      Environment & \multicolumn{2}{c}{PyTorch 1.10.0 with Python 3.8.12} \\
      Optimizer & \multicolumn{2}{c}{Adam (learning rate = 1e-4)} \\
      Computing platform & \multicolumn{2}{c}{NVIDIA GeForce GTX 3080Ti} \\
    \midrule 
      $\gamma$ in Equation \eqref{L_MDD} & \multicolumn{2}{c}{4} \\
      $\lambda$ in Equation \eqref{complete_L} & \multicolumn{2}{c}{1} \\
    \bottomrule
      \end{tabular}%
    \end{threeparttable}
    \begin{tablenotes} 
        \footnotesize 
        \item[1] $^{1}$ Training with only labeled source domain samples. 
    \end{tablenotes}
  \end{table}%

For comparison with referenced article \cite{zhang_variable-modulation_2022}, we employed Domain Adversarial Neural Network (DANN) \cite{bach_unsupervised_2015}. Simultaneously, within the realm of classical UDA models, we also employed Joint Adaptation Network (JAN) \cite{long_JAN_2017}, Batch Spectral Penalization (BSP) \cite{chen_BSP_2019}, and Minimum Class Confusion (MCC) \cite{jin_MCC_2020} for comparative experiments. As an exploratory endeavor, we also employed the Semi-Supervised Learning (SSL) model FixMatch \cite{sohn_Fixmatch_2020}. 

\subsection{Experiment Results and Analysis}\label{subsec5_3}

\begin{figure}[!b]
    \centering
    \vspace{-.3cm}
    \subfloat[DRSN]{\includegraphics[width=.48\columnwidth]{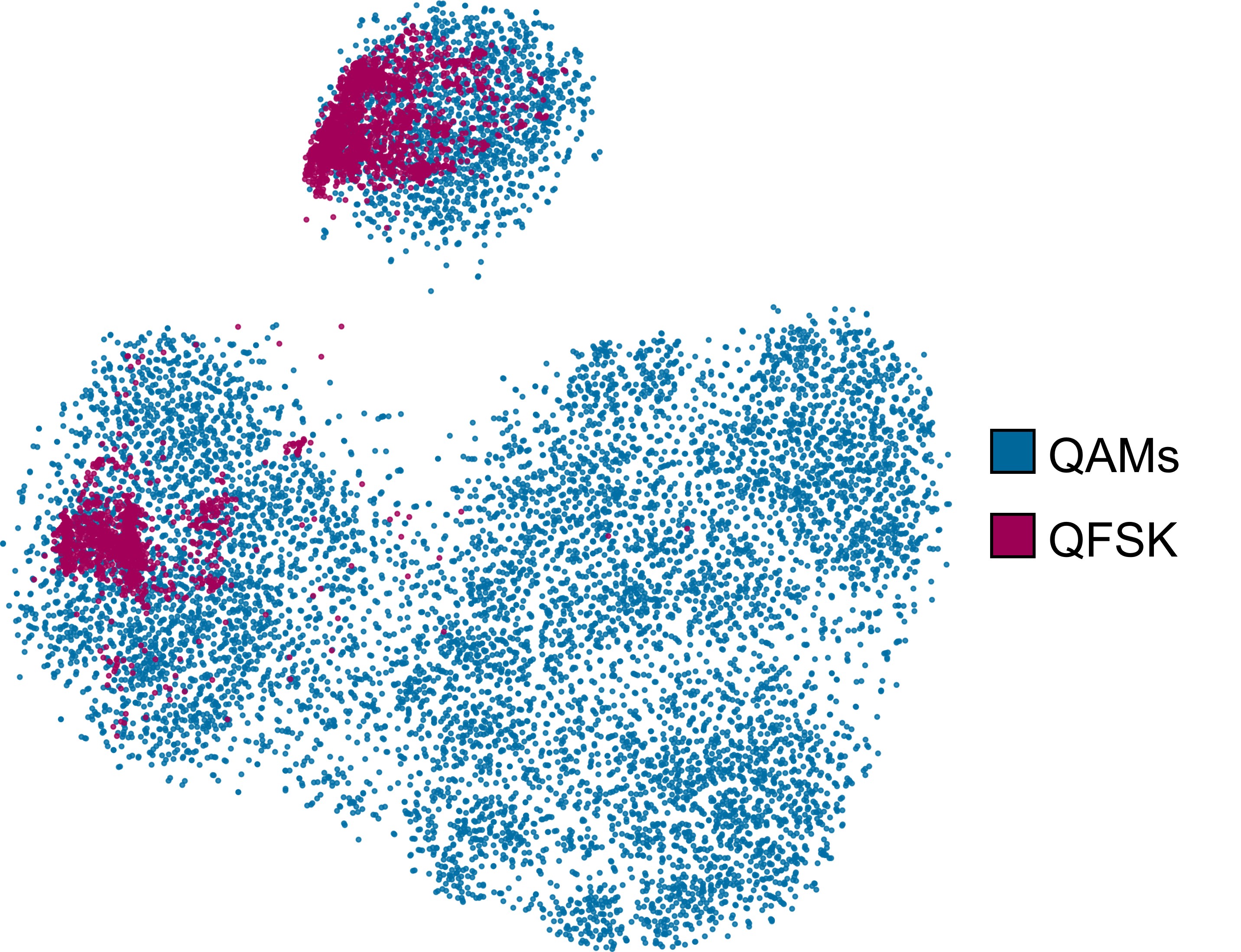}\label{dp1}}\hspace{0pt}
    \subfloat[DRSN with MDD]{\includegraphics[width=.48\columnwidth]{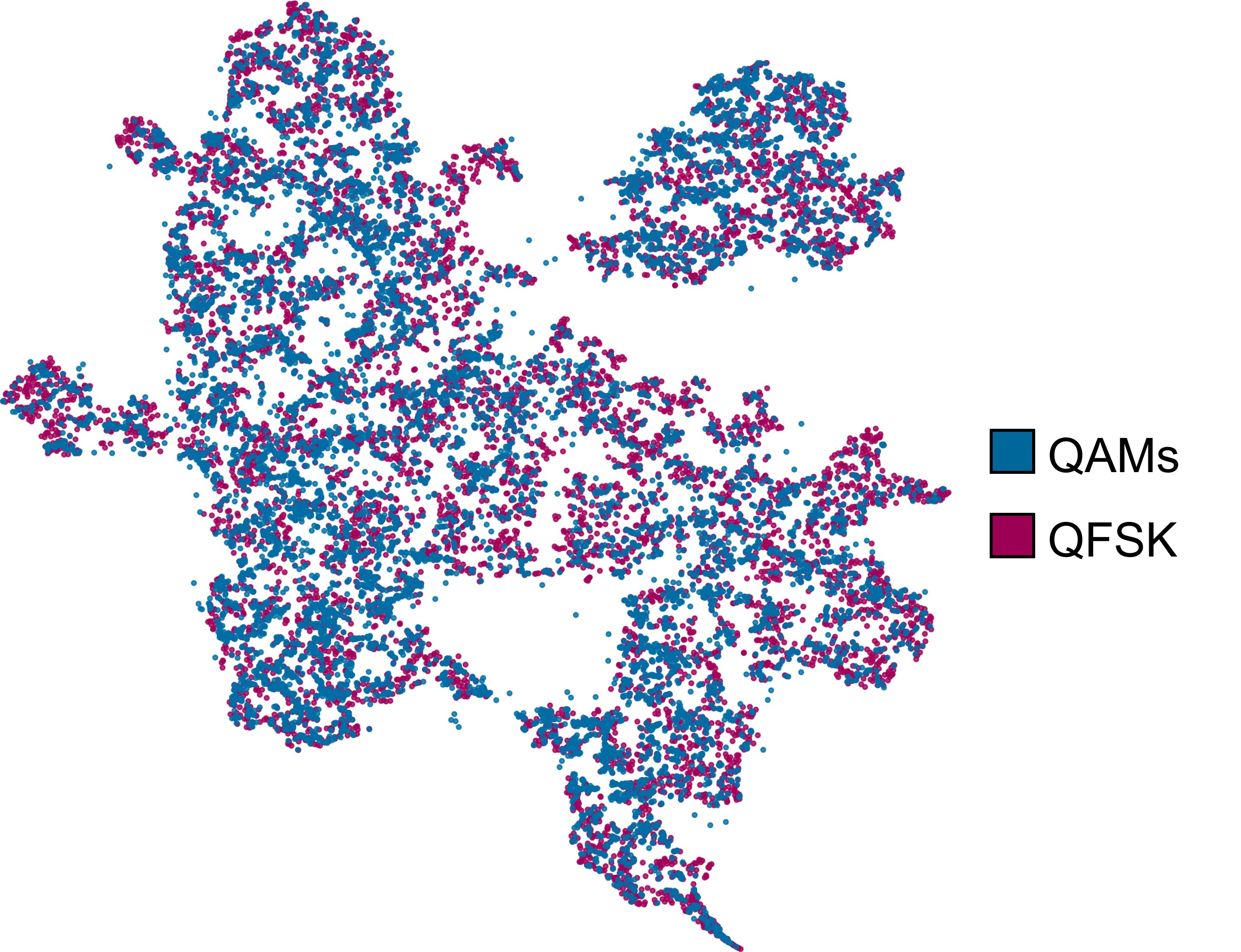}\label{dp2}}\\
    \vspace{-.2cm}

    \subfloat[DRSN on QAM]{\includegraphics[width=.48\columnwidth]{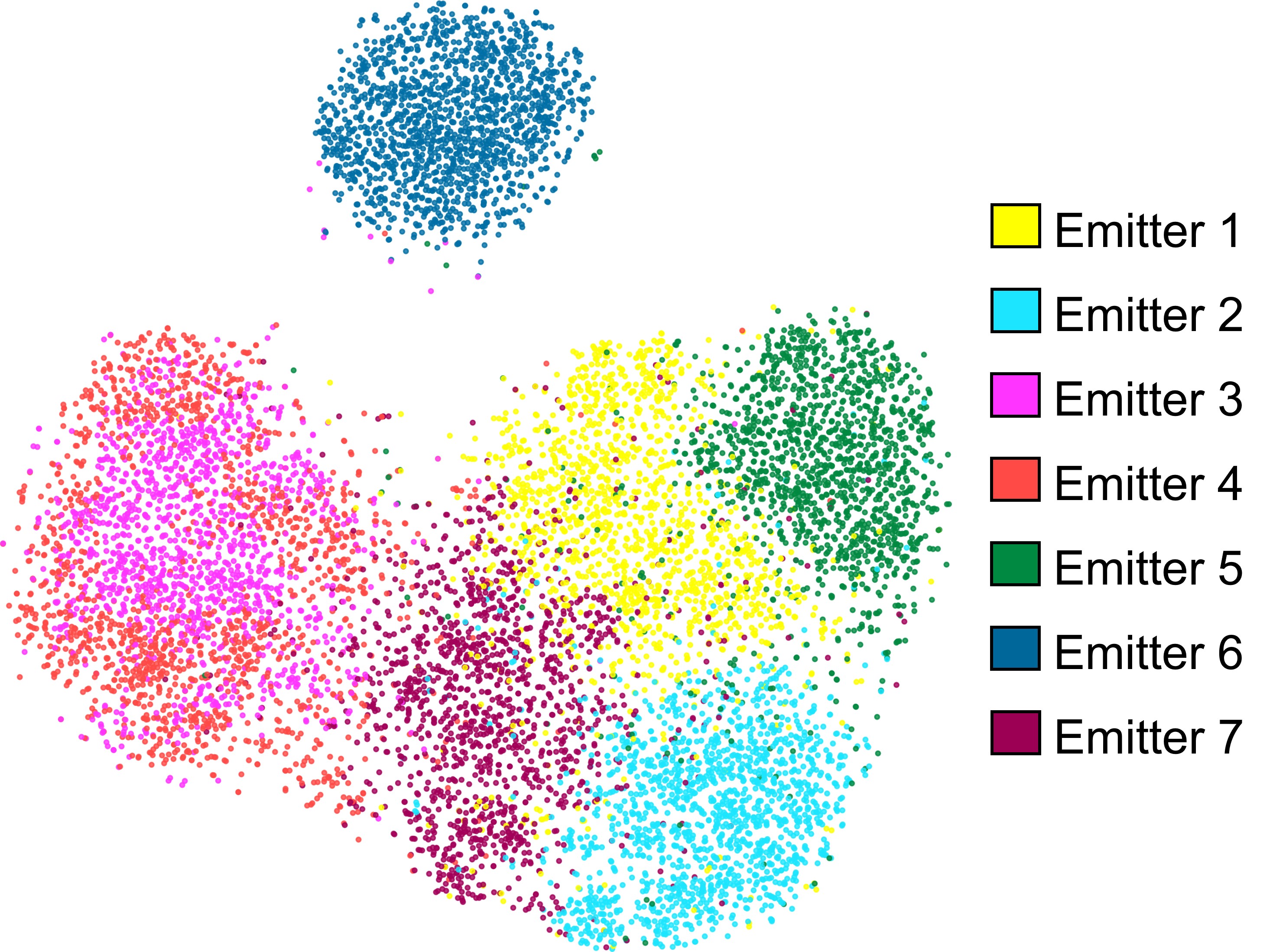}\label{dp3}}\hspace{0pt}
    \subfloat[DRSN on QAM with MDD]{\includegraphics[width=.48\columnwidth]{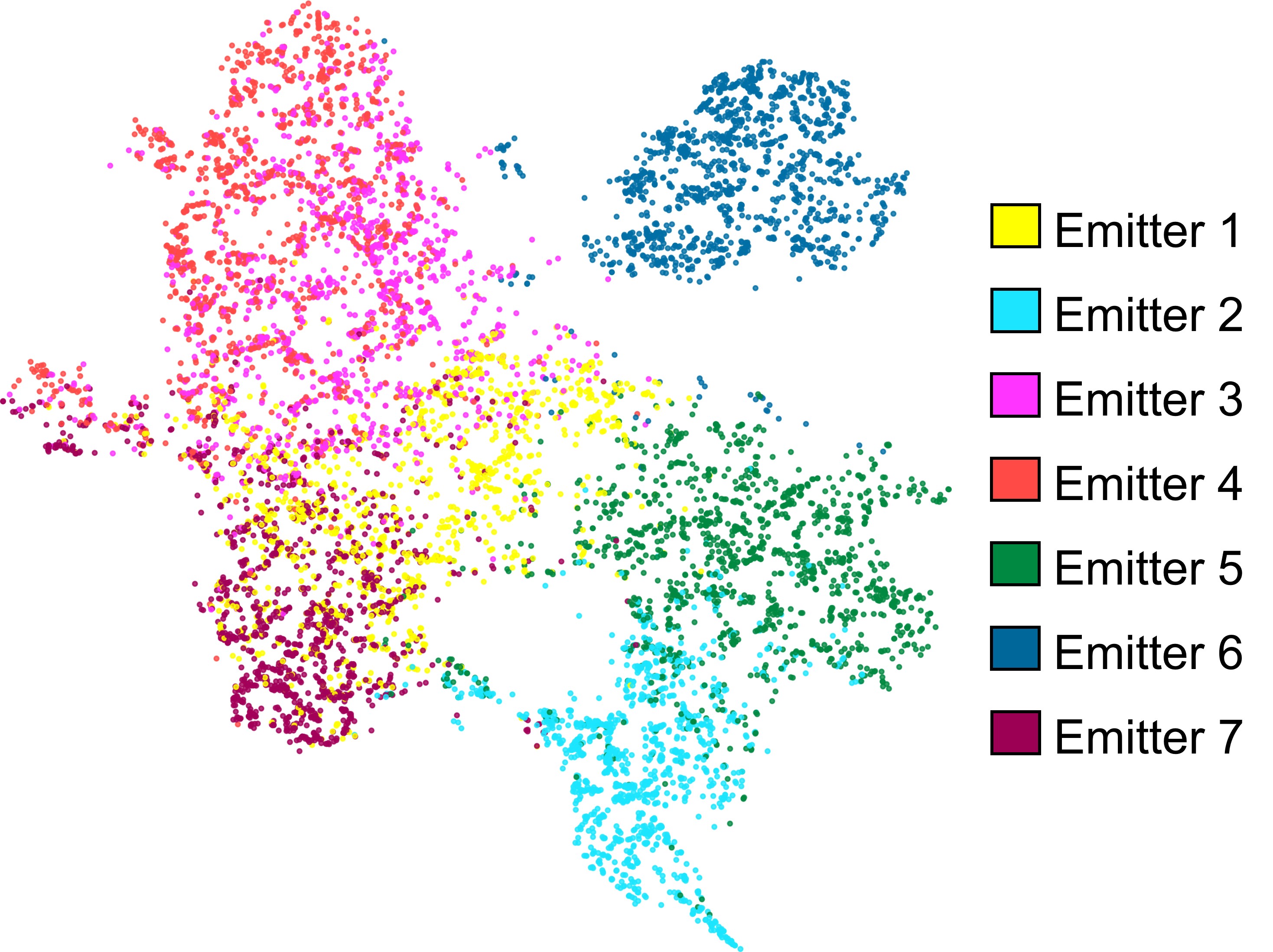}\label{dp4}}\\
    \vspace{-.2cm}

    \subfloat[DRSN on QFSK]{\includegraphics[width=.48\columnwidth]{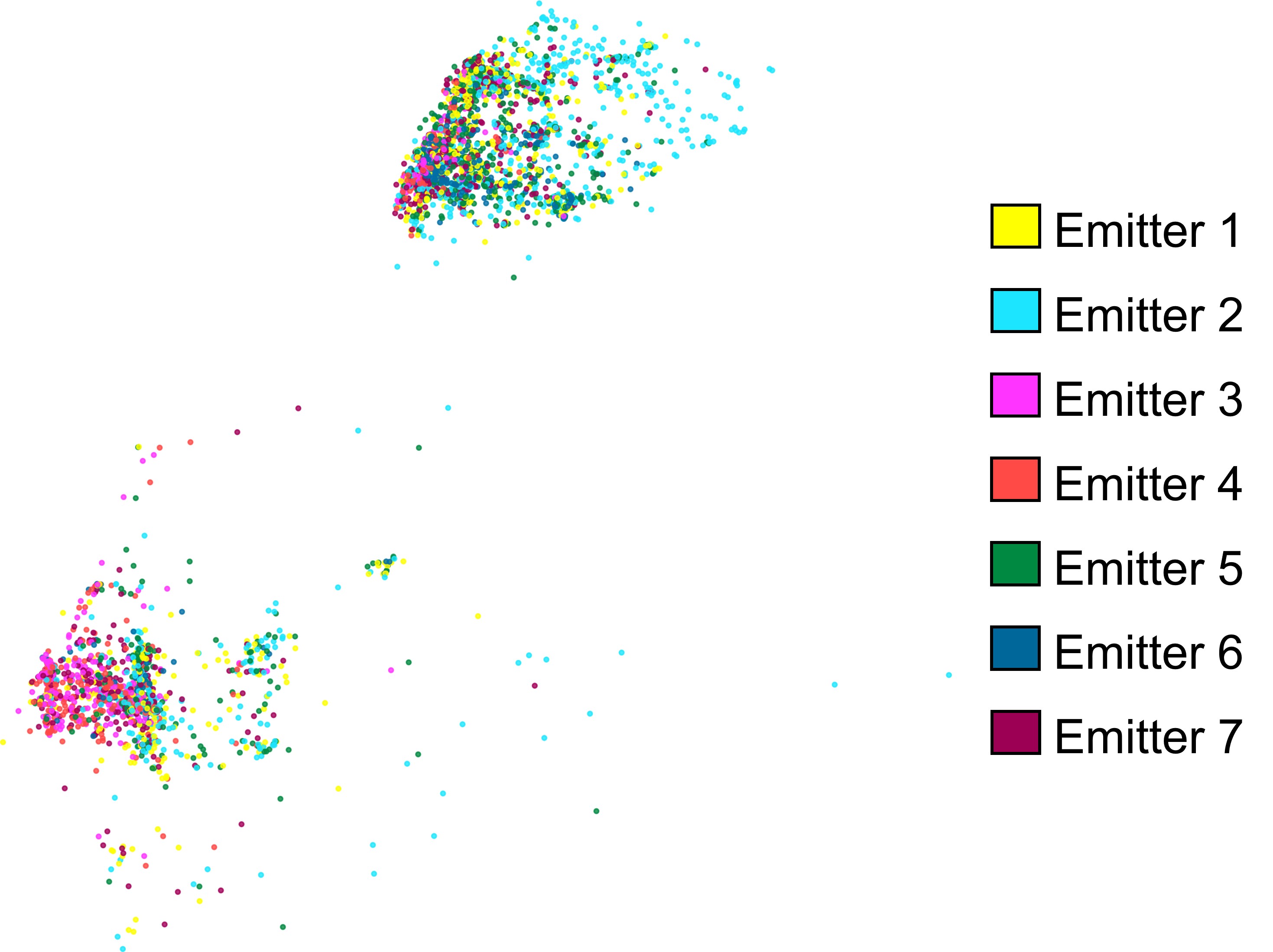}\label{dp5}}\hspace{0pt}
    \subfloat[DRSN on QFSK with MDD]{\includegraphics[width=.48\columnwidth]{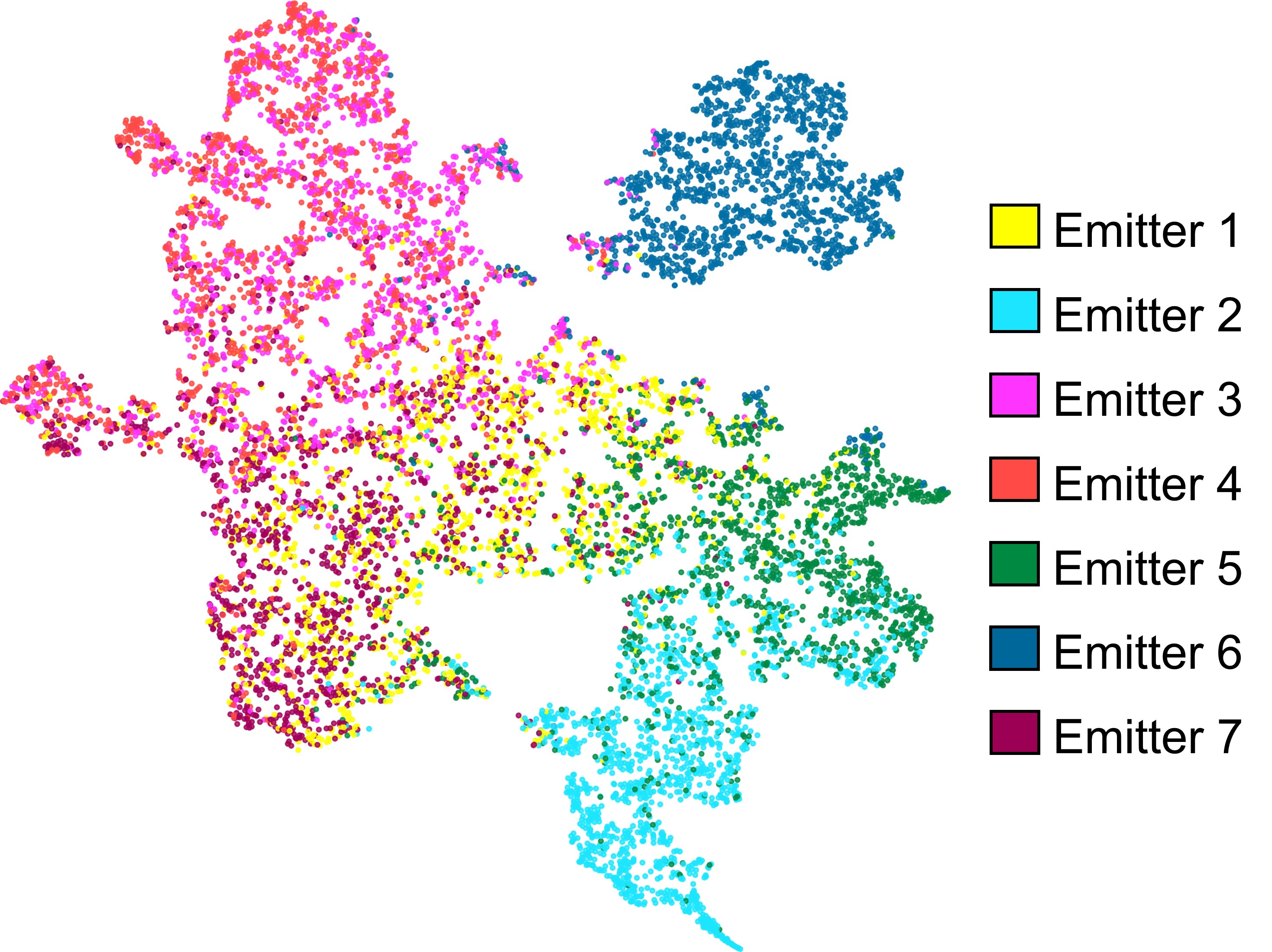}\label{dp6}}\\
    \caption{t-SNE graphs of extracted features on QAM and QFSK. (a) is a combination of (c) and (e), while (b) is a combination of (d) and (f).}\label{drsn_proposed}
\end{figure}

\begin{table*}[htbp]
    \centering
    \caption{Performance Comparison with SEI models}
    \setlength{\tabcolsep}{4.7mm}
\begin{threeparttable}
    \begin{tabular}{rrrrrrrrr}
    \toprule
    & Resnet with GAF & DRSN\tnote{1}  & DANN\tnote{2}  & JAN   & BSP   & MCC   & FixMatch & MDD \\
    \midrule
    {FM $\rightarrow$ Target}    & 25.25  & 33.65  & 53.92  & 54.41  & \underline{55.68}  & 44.68  & 45.35  & \textbf{55.86} \\
    {FSK$\rightarrow$ Target}   & 27.71  & 32.50  & 37.96  & \underline{38.86}  & 38.68  & 35.38  & 36.45  & \textbf{50.87} \\
    {PSK$\rightarrow$ Target}   & 33.42  & 34.87  & 46.20  & \underline{50.36}  & 48.62  & 37.88  & 43.37  & \textbf{54.13} \\
    {QAM$\rightarrow$ Target}   & 33.45  & 34.90  & 42.37  & \underline{48.02}  & 47.61  & 40.14  & 44.81  & \textbf{58.06} \\
    {FSK\_PSK$\rightarrow$ Target} & 24.93  & 24.33  & \underline{29.51}  & 27.71  & 28.71  & 25.46  & 23.44  & \textbf{50.49} \\
    \midrule
    \makebox[1.5cm][l]{\textbf{Average}}   & 28.95  & 32.05  & 41.99  & \underline{43.87}  & 43.86  & 36.71  & 38.68  & \textbf{53.88} \\
    \bottomrule
    \end{tabular}%
    \begin{tablenotes} 
        \footnotesize 
        \item[1] Subsequent models (DANN, JAN, BSP, MCC, FixMatch, and MDD) all employ DRSN with 18 layers as the residual network structure.
        \item[2] The implementation of DANN in SEI can be found in the referenced paper \cite{zhang_variable-modulation_2022}. 
    \end{tablenotes}
    \label{comparison}%
\end{threeparttable}
\end{table*}%

\begin{table*}[!t]
    \centering
    \begin{threeparttable}
    \caption{Performance Ablation with DRSN}
    \setlength{\tabcolsep}{3.5mm}
    \label{rsnet_dann_proposed}
    \begin{tabular}{l cc cc cc cc cc}
    \toprule
    \multirow{2}{*}{}&
\multicolumn{2}{c}{FM $\rightarrow$ Target}&\multicolumn{2}{c}{FSK $\rightarrow$ Target}&\multicolumn{2}{c}{PSK $\rightarrow$ Target}&\multicolumn{2}{c}{QAM $\rightarrow$ Target}&\multicolumn{2}{c}{FSK\_PSK $\rightarrow$ Target}\cr
    \cmidrule(lr){2-3} \cmidrule(lr){4-5}\cmidrule(lr){6-7}\cmidrule(lr){8-9} \cmidrule(lr){10-11} 
    & DRSN  & MDD   & DRSN  & MDD   & DRSN  & MDD   & DRSN  & MDD   & DRSN  & MDD \cr
    \midrule
    Source $\rightarrow$ LFM   & \underline{95.54}  & \underline{\textbf{97.03}} & 39.64 & \textbf{62.68} & 33.19 & \textbf{56.16} & 28.61 & \textbf{58.45} & 23.11  & \textbf{53.23} \\
    Source $\rightarrow$ NLFM  & \underline{95.78}  & \underline{\textbf{96.35}} & 37.66 & \textbf{46.25} & 33.75 & \textbf{55.10} & 31.94 & \textbf{58.49} & 27.00  & \textbf{48.87} \\
    
    \midrule
    Source $\rightarrow$ BFSK  & 56.54  & \textbf{66.99} & \underline{96.02}  & \underline{\textbf{97.41}} & 46.80  & \textbf{63.65} & 34.86 & \textbf{63.38} & 23.91  & \textbf{53.64} \\
    Source $\rightarrow$ QFSK  & 28.11  & \textbf{49.92} & \underline{96.68}  & \underline{\textbf{97.70}} & 24.84 & \textbf{42.87} & 16.59 & \textbf{51.74} & 16.93  & \textbf{45.49} \\
    
    \midrule
    Source $\rightarrow$ BPSK  & 33.06  & \textbf{62.66} & 39.97 & \textbf{58.81} & \underline{94.50}  & \underline{\textbf{95.94}} & 50.36 & \textbf{61.89} & 32.09  & \textbf{52.99} \\
    Source $\rightarrow$ QPSK & 19.69  & \textbf{50.14} & 23.14 & \textbf{41.52} & \underline{94.85}  & \underline{\textbf{95.12}} & 34.16 & \textbf{49.12} & 23.57  & \textbf{46.76} \\
    Source $\rightarrow$ 8PSK  & 39.39  & \textbf{62.86} & 39.65 & \textbf{56.55} & \underline{95.18}  & \underline{\textbf{96.33}} & 32.03 & \textbf{63.34} & 26.59  & \textbf{52.21} \\
    
    \midrule
    Source $\rightarrow$ 16QAM & 38.08  & \textbf{48.03} & 31.08 & \textbf{38.10} & 31.04 & \textbf{49.10} & \underline{95.75}  & \underline{\textbf{95.93}} & 22.81  & \textbf{49.33} \\
    Source $\rightarrow$ 64QAM & 37.95  & \textbf{44.81} & 28.32 & \textbf{41.82} & 30.56 & \textbf{46.39} & \underline{94.39}  & \underline{\textbf{95.79}} & 22.95  & \textbf{51.93} \\
    
    \midrule
    Source $\rightarrow$ BFSK\_QPSK & 26.40  & \textbf{58.46} & 26.18 & \textbf{61.95} & 36.06 & \textbf{61.90} & 47.09 & \textbf{60.28} & \underline{98.68} & \underline{\textbf{98.71}} \\
    Source $\rightarrow$ QFSK\_BPSK & 23.65  & \textbf{58.86} & 26.90  & \textbf{50.16} & 42.75 & \textbf{57.88} & 38.47 & \textbf{55.87} & \underline{98.42} & \underline{\textbf{98.81}} \\
\midrule
    \makebox[1.8cm][r]{\textbf{Average}\tnote{1}} &  33.65  & \textbf{55.86} & 32.50  & \textbf{50.87} & 34.87  & \textbf{54.13} & 34.90  & \textbf{58.06} & 24.33  & \textbf{50.49} \\
    \bottomrule
    \end{tabular}
    \label{ablation}
    \begin{tablenotes} 
        \footnotesize 
        \item[1] The results are averaged over modulation scheme variations. This implies the values underlined on the diagonal are not included.
    \end{tablenotes}
    \end{threeparttable}
\end{table*}

We employed Residual Neural Network with Gramian Angular Field Image (Resnet with GAF) \cite{ma_gafrsnet_2023} and Deep Residual Shrinkage Network (DRSN) \cite{zhao_deep_2020} as baseline. 

The results on the 5 groups across 11 modulation schemes are presented in Table \ref{comparison}. The baselines, lacking the ability to utilize unlabeled samples during the training phase, exhibit average accuracies of only 28.95\% and 32.05\%, respectively, when confronted with modulation scheme transfer. Among models with unsupervised learning capability, our proposed MDD achieved the highest average accuracy of 53.88\%, surpassing the second-best method JAN by 10.01\%.

As an ablation, we removed the introduced MDD framework (depicted in the purple part in Fig. \ref{MDD_structure}), leaving only a traditional classifier with DRSN acting as the feature extractor. The results are presented in Table \ref{rsnet_dann_proposed}. Raw DRSN showcased exceptional SEI performance in the fixed modulation scheme. However, when faced with scenarios involving variable modulations, a substantial decrease was observed. In contrast, the proposed MDD framework demonstrated significant improvement. Compared to training directly without incorporating unlabeled samples, our approach achieved an average improvement of 21.83\% over DRSN. 

Following the introduction of the proposed MDD method, the model effectively addresses the issue described in Fig. \ref{decisions}, aligning the features of different modulation schemes and, to some extent disregarding the modulation dimension. To evaluate the performance, we employed t-Distributed Stochastic Neighbor Embedding (t-SNE) to visualize the output of the feature extractor, observing the performance of raw DRSN and MDD on labeled (QAM) and unlabeled (QFSK) modulations, as depicted in Fig. \ref{drsn_proposed}.

\begin{figure}[htbp]
    \centering
    \subfloat[DRSN on QAM]{\includegraphics[width=.5\columnwidth]{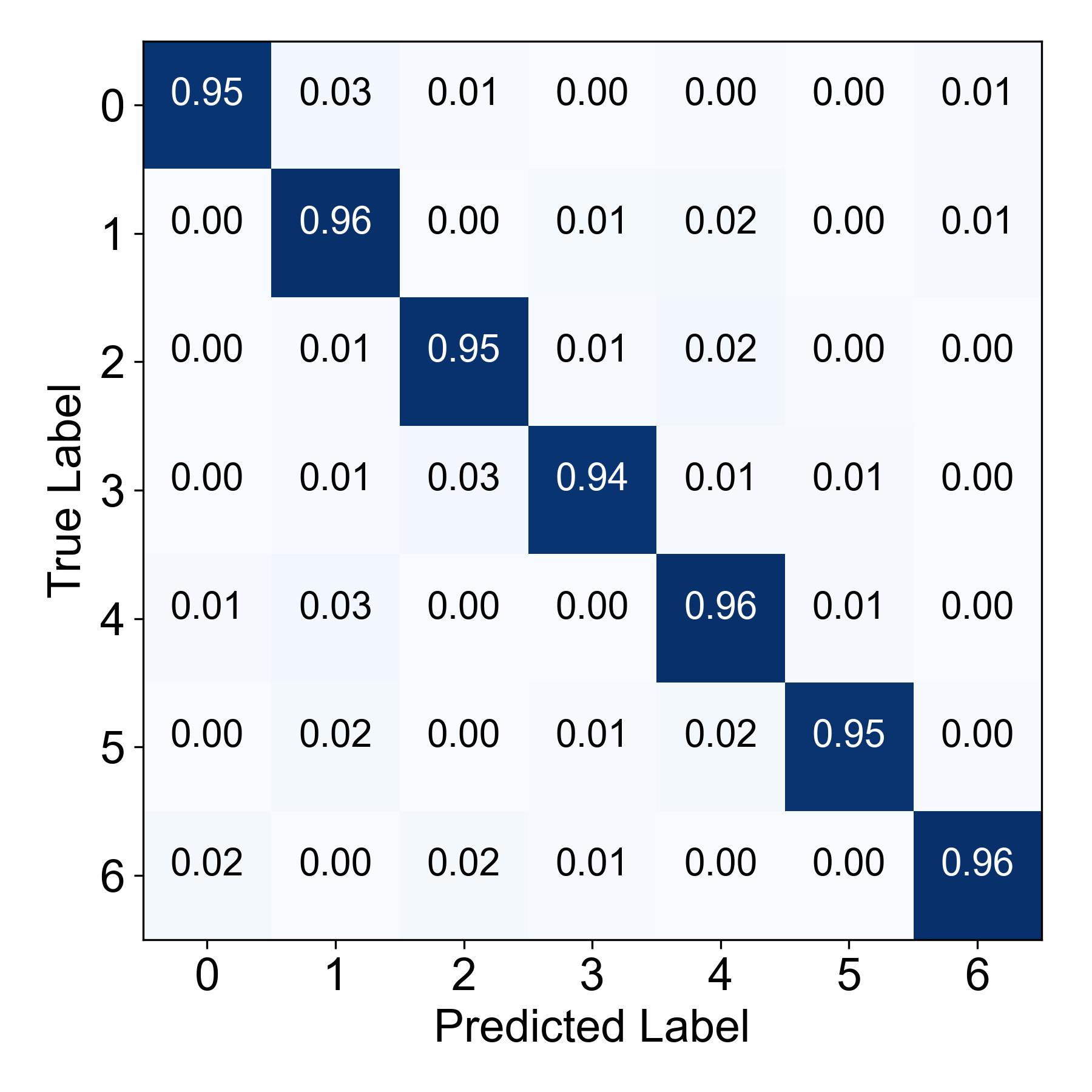}\label{confmat_nice_origin}}
    \subfloat[DRSN on QAM with MDD]{\includegraphics[width=.5\columnwidth]{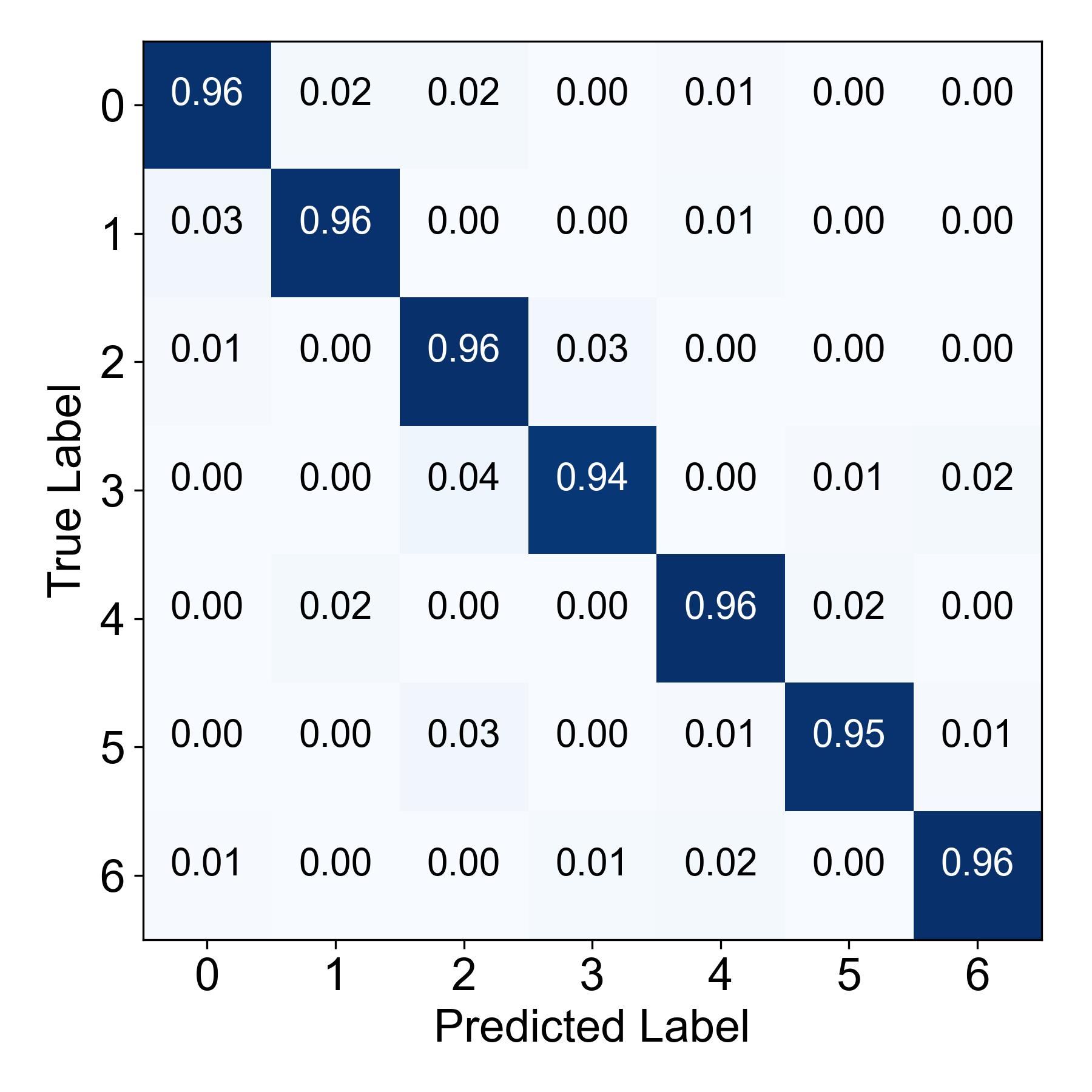}\label{confmat_nice}}\\
    \vspace{-.3cm}
    \subfloat[DRSN on QFSK]{\includegraphics[width=.5\columnwidth]{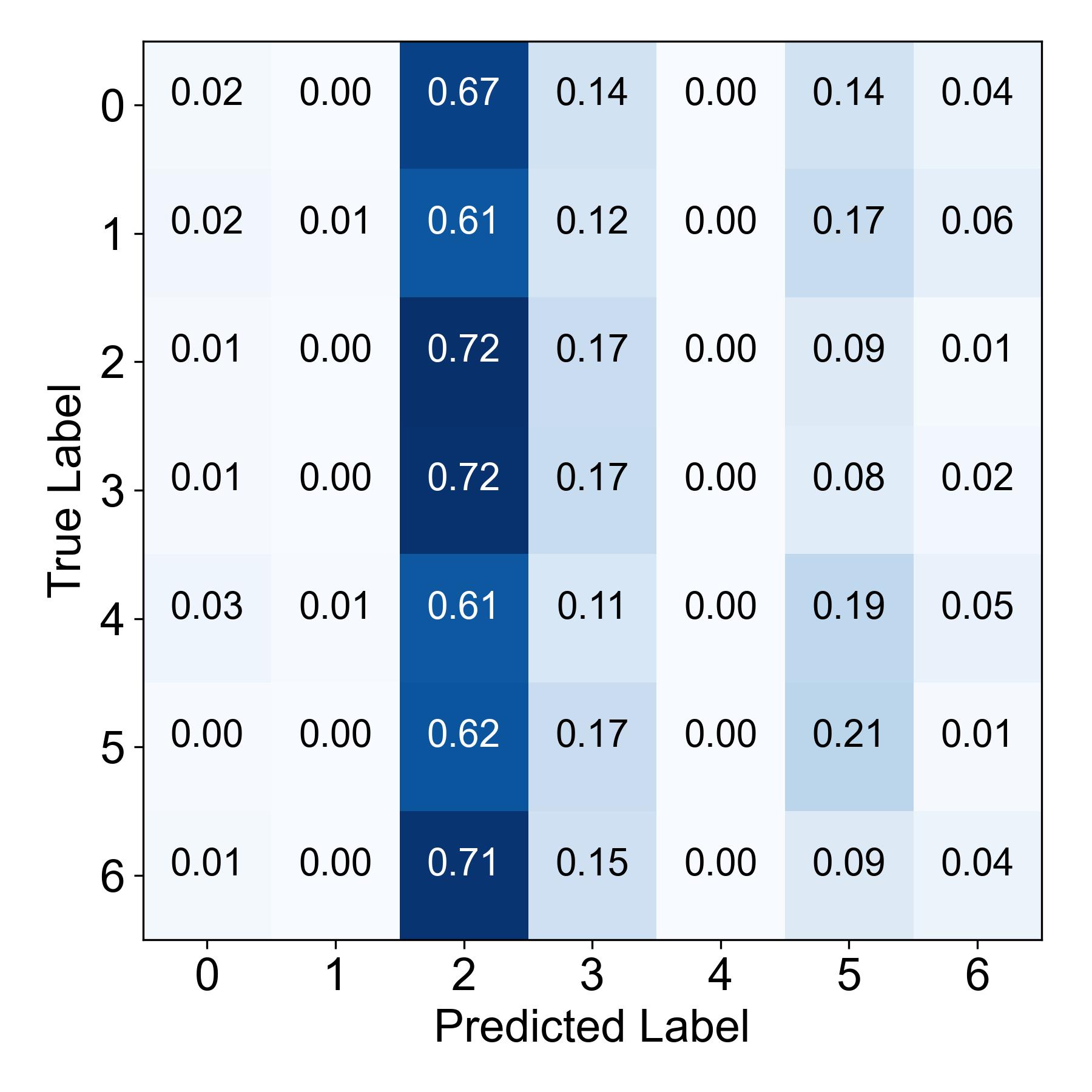}\label{confmat_bad}}
    \subfloat[DRSN on QFSK with MDD]{\includegraphics[width=.5\columnwidth]{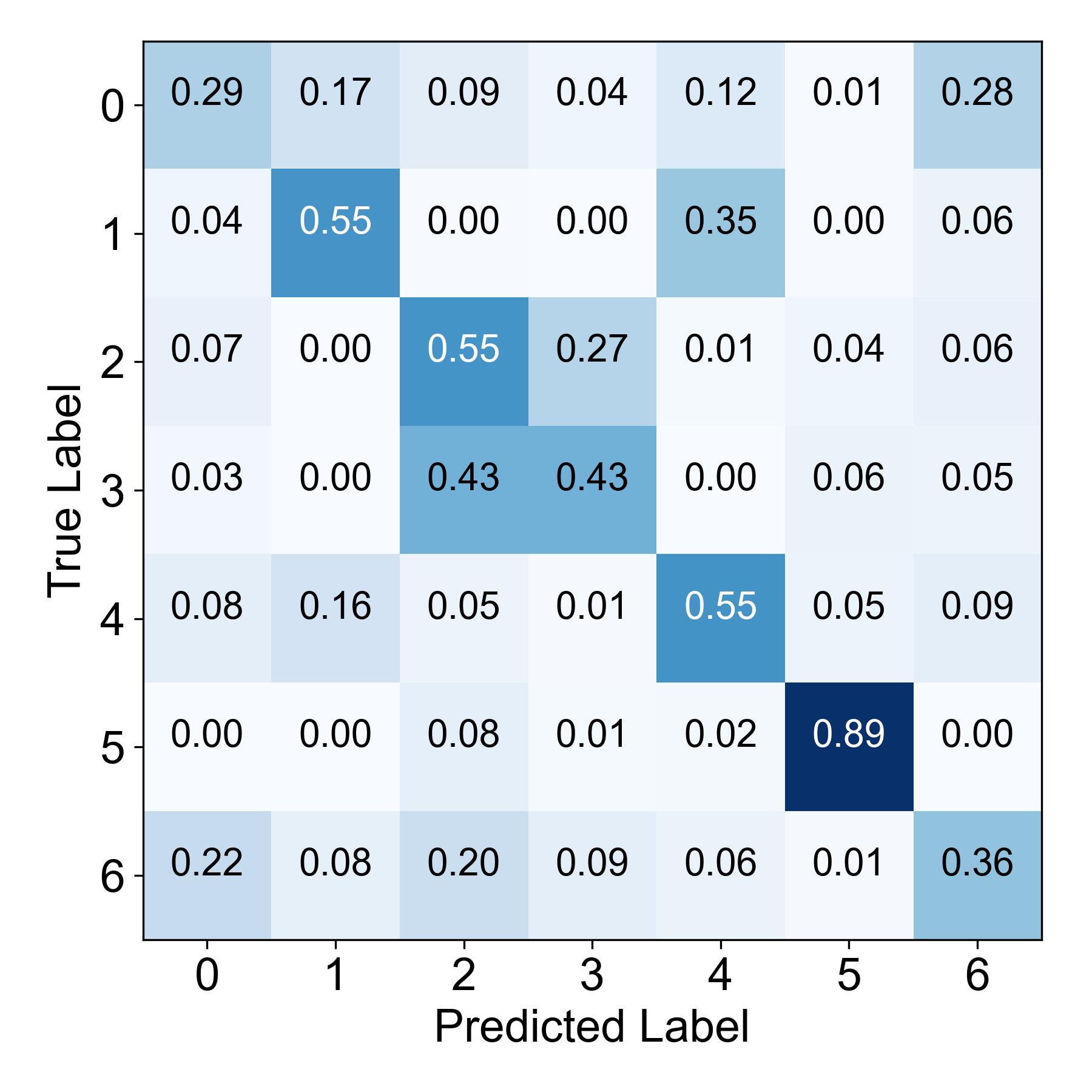}\label{confmat_good}}
    \caption{The confusion matrix of all transmitters on QAM and QFSK.}\label{confmat}
    \vspace{-.3cm}
\end{figure}

In comparison to the t-SNE graph depicted in Fig. \ref{dp1}, where DRSN distinctly separates the features of QAM and QFSK, the proposed method aligns the features of QAM and QFSK, as shown in Fig. \ref{dp2}. Both models effectively classified the sources emitted QAM signals, as demonstrated in Fig. \ref{dp3} and \ref{dp4}. However, when it comes to classifying QFSK, the proposed MDD exhibits an equivalent capability to classifying QAM, while DRSN completely fails in this task. As indicated in the QAM $\rightarrow$ BPSK entry in Table \ref{comparison}, the average accuracy increased from 16.59\% to 51.74\%. The confusion matrix of all transmitters is presented in Fig. \ref{confmat}. Under a fixed modulation scheme, both DRSN and MDD demonstrate outstanding performance. However, when facing modulation variations, DRSN nearly categorizes all transmitters into a single class. MDD effectively alleviates this issue.

Through Table \ref{cost}, we can observe the memory consumption and training time for different methods. Except for MDD, other methods only introduce a loss function without adding model parameters, resulting in the same space occupancy. The proposed MDD introduces an additional classifier, but its time consumption is only about 1.25 times that of the baseline DRSN. In comparison to other methods, MDD also exhibits relatively lower time consumption.

\begin{table}[!t]
    \centering
\begin{threeparttable}
    \caption{Computational Efficiency of Different Methods}
    \begin{tabular}{lrrr}
    \toprule
        Model & \multicolumn{1}{l}{\textbf{FLOPs / M}} & \multicolumn{1}{l}{\textbf{ Memory Cost / Mb}} & \multicolumn{1}{l}{\textbf{Time Cost\tnote{1} / s}} \\
    \midrule
        DRSN  & 572.48  & 13.29  & 1.53  \\
    \midrule
    FixMatch & 715.87  & 20.20  & 3.63  \\
    MCC   & 724.89  & 20.20  & 3.20  \\
    BSP   & 721.88  & 20.20  & 2.97  \\
    JAN   & 718.88  & 20.20  & 2.93  \\
    MDD  & 728.43  & 20.73  & \underline{1.91}  \\
    DANN  & 715.87  & 20.20  & \textbf{1.70}  \\
    \bottomrule
        \end{tabular}%
    \label{cost}
    \begin{tablenotes} 
        \footnotesize 
        \item[1] The average time consumed per batch with a batch size of 32 for 1000 batches.
    \end{tablenotes}
\end{threeparttable}
\end{table}

\section{Conclusion}\label{sec5}

In this research, we propose an SEI method tailored to scenarios involving transmitters with varying modulation schemes. We address the issue of SEI classifiers prioritizing modulation scheme over RFF features by introducing an SEI framework with MDD. Experimental results across variable modulation schemes demonstrate that our proposed method effectively aligns modulation features while emphasizing RFF characteristics. In comparison to classical SEI models and UDA approaches, our method consistently achieves superior accuracy. Future work entails devising more effective strategies for decoupling modulation schemes and RFF features, as well as enhancing recognition accuracy under diverse parameter variation scenarios.

\section*{Acknowledgement}
This work was supported by Natural Science Foundation of China (61906038 and 62006119).

\bibliographystyle{IEEEtran}
\bibliography{ref}

\end{document}